\documentstyle[aps,preprint]{revtex}

\begin{document}

\tighten

\title{Covariant four-dimensional scattering equations for the $NN-\pi
NN$ system}

\author{D.~R.~Phillips and I.~R.~Afnan}

\address{Department of Physics, The Flinders University of South Australia,\\
GPO Box 2100, Adelaide 5001, Australia.}

\maketitle

\begin{abstract}
We derive a set of coupled four-dimensional integral equations for the
$NN-\pi NN$ system using our modified version of the Taylor method of
classification-of-diagrams. These equations are covariant, obey two and
three-body unitarity and contain subtraction terms which eliminate the
double-counting present in some previous four-dimensional $NN-\pi NN$
equations. The equations are then recast into a from convenient for
computation by grouping the subtraction terms together and obtaining a set of
two-fragment scattering equations for the amplitudes of interest.
\end{abstract}

\newtheorem {theorem}{Theorem}
\newtheorem {defn}{Definition}
\newtheorem {lemma}{Lemma}
\newtheorem {claim}{Claim}
\newcounter {class}

\newcommand{\be}{\begin {equation}}
\newcommand{\ee}{\end {equation}}
\newcommand{\bea}{\begin {eqnarray}}
\newcommand{\eea}{\end {eqnarray}}
\newcommand{\bel}{\begin {eqnarray*}}
\newcommand{\eel}{\end {eqnarray*}}
\newcommand{\pifact}{-\frac{1}{2 \pi i}}
\newcommand{\fadj}{f^\dagger}
\newcommand{\go}[1]{g_{N_{#1}}}
\newcommand{\gpo}[1]{g_{\pi N_{#1}}}
\newcommand{\gtb}{d_1 d_2 d_\pi}
\newcommand{\g}[1]{g_0^{#1}} \newcommand{\gu}{g_0}
\newcommand{\implies}{\Rightarrow}
\newcommand{\nn}{\nonumber}
\newcommand{\bpsi}{\overline{\psi}}
\newcommand{\ub}{\overline{u}}
\newcommand{\vb}{\overline{v}}
\newcommand{\adja}{a^{\dagger}}
\newcommand{\ba}{b^{\dagger}}
\newcommand{\da}{d^{\dagger}}
\newcommand{\phia}{\phi^{\dagger}}
\newcommand{\half}{\frac{1}{2}}
\newcommand{\A}[4]{A_{#2 \leftarrow #1}^{(#4,#3)}}
\newcommand{\At}[2]{A_{#2 \leftarrow #1}}
\newcommand{\fir}[1]{f^{(#1)}}
\newcommand{\fai}[1]{{f^{(#1)}}^{\dagger}}
\newcommand{\fsai}[2]{{f^{(#1)}_{#2}}^{\dagger}}
\newcommand{\tfai}[1]{\tilde{f}^{(#1) \dagger}}
\newcommand{\tpni}[1]{t_{\pi N}^{(#1)}}
\newcommand{\tit}[1]{\tilde{t}^{(#1)}}
\newcommand{\tait}[1]{\tilde{t}^{(#1) \dagger}}
\newcommand{\ttpni}[1]{\tilde{t}_{\pi N}^{(#1)}}
\newcommand{\ttpna}[1]{\tilde{t}_{\pi N}^{(#1) \dagger}}
\newcommand{\tnni}[1]{T_{N N}^{(#1)}}
\newcommand{\titnni}[1]{\tilde{T}_{N N}^{(#1)}}
\newcommand{\bfi}[1]{F^{(#1)}}
\newcommand{\bfai}[1]{{F^{(#1)}}^{\dagger}}
\newcommand{\cfo}[2]{C_4^{\{#1\}\{#2\}}}
\newcommand{\cfi}[1]{C_5^{\{#1\}}}
\newcommand{\bd}[2]{\bar{\delta}_{#1 #2}}
\newcommand{\titit}[1]{\tilde{\tilde{t}}^{(#1)}}

\section {Introduction}

\label {sec-intro}

The $NN-\pi NN$ problem occupies a privileged place in nuclear
physics. Not only does its history stretch back to Yukawa's
original efforts to model the strong nuclear force \cite{Yu39},
but it is still the subject of considerable research to this
day. This current interest occurs partly because the $NN-\pi
NN$ system is one light nuclear system which can be used to
test the validity of models in which hadrons are the
fundamental degrees of freedom. Only in systems which, like
the $NN-\pi NN$ system, contain a relatively small number of
degrees of freedom, is it possible to complete a calculation
of experimentally observed quantities within the framework of such a
model, while retaining some control over the approximations used.
Consequently, considerable effort has been put into attempting to produce
physically accurate hadronic models of the $NN-\pi NN$ system. The
predictions of these models can then be compared with the mass
of available experimental data and a judgement on the models'
validity formed.

Over the past twenty years considerable theoretical progress
has been made in  this direction, with the culmination being
the independent derivation, by a number of groups using
different techniques,  of a set of scattering equations known
as the
$NN-\pi NN$
equations~\cite{Th73,Mi76,MK77,Ri77,TR79,AM79,AM80,AM81,AM83,AB80,AB81}.
The problem is that models based on these $NN-\pi NN$ equations
which treat the $\pi-N$ amplitude as the sum of a pole and non-pole
term fail to reproduce the experimental data. This may well
be because of certain theoretical inconsistencies in the equations. For
example, Jennings \cite{Je88} has pointed out that the $NN-\pi NN$
equations include the diagram on the right of Figure \ref{fig-Jennings},
but exclude that on the left---even though the left-hand diagram is
merely a different time-order of the right-hand one and both diagrams
represent the same physical process. Jennings and Rinat
\cite{Je88,JR88} and Mizutani et al. \cite{Mi88} have made it
plausible that the omission of this diagram accounts for the failure of
models based on the $NN-\pi NN$ equations to correctly predict the tensor
polarization, $T_{20}$ for $\pi-d$ scattering. (For more detail
on the history of the theory of the $NN-\pi NN$ system and a
thorough comparison with the experimental data, see the recent
book by Garcilazo and Mizutani \cite{GM90}.)

The standard $NN-\pi NN$ equations do not include the Jennings mechanism
because they are derived by using unitarity as a criterion for
truncating the full field theory of nucleons and pions. One way in which
this derivation can be done is to examine the diagrammatic expansion for
the $NN-\pi NN$ amplitudes in old-fashioned or time-ordered
perturbation theory (TOPT). The $NN-\pi NN$ equations may then be
obtained by truncating this expansion at the one explicit pion level.
(See \cite{AB80,AB85} for details.) This truncation not only appears to
lead to incorrect predictions for $T_{20}$, but also produces the
under-dressing of the two-nucleon propagator discussed in
\cite{Sa85,BK92A}.  Indeed, the fundamental shortcoming of the standard
$NN-\pi NN$ equations is that two diagrams each representing the
same set of physical processes, but having them occur in a different
time-order may end up treated completely differently. In fact, as we see
in Fig.~\ref{fig-Jennings}, one may be included and one excluded, with
potentially disastrous consequences.

This difficulty can be entirely circumvented, however, if the
diagrammatic expansion of the field theory of nucleons and pions is
performed in covariant perturbation theory, rather than time-ordered
perturbation theory. This is so because TOPT diagrams which differ only
in the time-order of the physical interactions involved are all
included in the one covariant perturbation theory graph. In particular,
the two TOPT diagrams in Figure~\ref{fig-Jennings} arise from the same
covariant perturbation theory graph. Hence, if equations for the
$NN-\pi NN$ system are derived from this covariant perturbation theory
diagrammatic expansion they will not suffer from the fundamental
deficiency of the standard $NN-\pi NN$ equations.

Pursuing this approach to the $NN-\pi NN$ system also has the advantage
that, provided care is taken in truncating the field theory, the
equations will automatically be covariant, since they are derived from
a covariant diagrammatic expansion. Furthermore, if integral equations
are used to sum classes of diagrams containing infinitely many
perturbation graphs, they generate amplitudes which are {\em
non}-perturbative, even though the equations themselves may originally
be derived from the perturbative diagrammatic expansion.
Examples of this are the Schwinger-Dyson equations of a field theory
which may be derived from a resummation of the original
perturbation expansion of the theory, but in no way involve
a truncation of this expansion at some order in the coupling constant of
the theory. Such integral equations governing the $NN-\pi NN$ system
will of necessity be four-dimensional, and hence the numerical solution
of these equations in order to obtain experimental predictions is a
challenging problem. However, until this problem is tackled it is not
clear that the one-explicit-pion sector of the $NN-\pi NN$ system has
been properly dealt with. Indeed, the standard theory of the $NN-\pi NN$
system, based on time-ordered perturbation theory, only approximates the
fuller description based on a covariant diagrammatic expansion. Until it
is clear that this approximation is an adequate one appeals to mechanisms
beyond the one-pion sector to remedy the present disagreement between
theory and experiment cannot be definitely upheld or overturned. The
derivation of equations for the $NN-\pi NN$ system in the framework of
covariant perturbation  theory is therefore an important question at the
heart of one of the oldest problems in theoretical nuclear physics.

The question then is: how are such equations to be derived? Given a
Lagrangian one could use functional techniques to derive the relevant
Schwinger-Dyson equations of the field theory, truncating the
Schwinger-Dyson equation hierarchy by some approximation scheme, and so
obtaining a set of four-dimensional coupled integral equations for the
$NN-\pi NN-\pi \pi NN-\ldots$ system. In this paper we choose not to
employ such a functional calculus technique, but instead use the Taylor
method of classification-of-diagrams \cite{Ta63}.

The Taylor method is a general one allowing the derivation of
equations connecting the amplitudes obtained from diagrammatic
expansions, and hence is ideally suited to the derivation of equations
from the Feynman diagrammatic expansion of the amplitudes for the
$NN-\pi NN$ system. Reference  \cite{AP94}, henceforward known as
paper I, presented a review of the Taylor method. That paper also
pointed out that Taylor's method leads to double-counting  when
applied to certain covariant perturbation theory amplitudes.
It was shown how the Taylor method could be modified in order
to eliminate this double-counting, thus producing a technique
by which double-counting-free covariant four-dimensional scattering
equations may be derived for, not only the $NN-\pi NN$ system, but also
other few-hadron systems, such as the $\pi N-\pi \pi N$ system and the
problem of pion photoproduction.

This idea of applying Taylor's original method to the $NN-\pi NN$ system
is not a new one. Three pairs of authors have already attempted to derive
covariant four-dimensional equations for the $NN-\pi NN$ system using Taylor's
method. Firstly, Avishai and Mizutani (AM), derived coupled covariant
four-dimensional equations for the $NN-\pi NN$ system using the Taylor
method~\cite{AM83}. However, Avishai and Mizutani failed to eliminate the
double-counting which arises when the Taylor method is applied directly to
the graphical expansion of a time-dependent perturbation theory.
Consequently the equations they derived double-counted certain diagrams.
Avishai and Mizutani
then used Blankenbecler-Sugar \cite{BbS66} reduction in order to reduce
their covariant four-dimensional integral equations to more manageable
three-dimensional ones \cite{La87}. The reduced equations thus obtained are
equivalent to the standard three-dimensional $NN-\pi NN$ equations AM (and
others) had previously derived. The standard $NN-\pi NN$ equations do not
contain any double-counting  and thus the final set of equations used for
numerical work by AM may be regarded as double-counting free. However, AM's
``derivation" of these equations by a three-dimensional reduction of
four-dimensional equations which themselves {\em do} contain double-counting is
open to question.

Secondly, in 1985 Afnan and Blankleider (AB) derived a set of
covariant $BB-\pi BB$ equations, in which the baryon $B$ could
be either a nucleon or a delta \cite{AB85}. However, instead
of using Taylor's original classification-of-diagrams scheme
Afnan and Blankleider used a simplification of Taylor's
method which they, together with Thomas and Rinat, had
developed some years before \cite{TR79,AB80}. (This
simplification and its relation to Taylor's  original work were
discussed in paper I\@.) The equations thus obtained by Afnan
and Blankleider were exactly those found by Avishai and
Mizutani, except that the nucleon $N$ in Avishai and
Mizutani's theory was replaced throughout by a baryon
$B$ which could be either a nucleon or a delta \footnote{There were terms
present in AM's equations which were not included   in AB's results, but the
addition of these terms to AB's equations could have been effected with only
minor changes to their argument.}. The use of a diagrammatic technique
similar to that used by Avishai and Mizutani, and the consequent derivation
of similar equations naturally meant that Afnan and Blankleider's equations
also contained double-counting if viewed in a four-dimensional framework.
However, as in Avishai and Mizutani's case, this problem was never fully
revealed, since the
$BB-\pi BB$ equations were solved in a three-dimensional time-ordered
perturbation theory, and so the double-counting was temporarily hidden, even
though it was still present in the full four-dimensional theory.

Recently, Kvinikhidze and Blankleider (KB) have recognized the
double-counting
in these two derivations \cite{BK94A,BK94B}. They have introduced a
modification to Taylor's method as applied by AM and AB which allows them to
derive a set of covariant four-dimensional equations for the $NN-\pi NN$
system
which are free from double-counting. These equations are, apart from a minor
point, equivalent to the ones derived here. However, this work differs from
that
of Ref.~\cite{BK94A,BK94B} in three main ways. Firstly, the method used here is
rigorously based on the modification of Taylor's original classification scheme
for an $m \rightarrow n$ amplitude in any perturbation theory as detailed in I.
By contrast, KB have used a classification-of-diagrams scheme which is only
loosely based on Taylor's original work. Secondly, our use of the full Taylor
method allows us to exploit the true Lagrangian independence of that technique.
Hence here we do not specify the Lagrangian to be used in the description of
the
$NN-\pi NN$ system, while KB restricted themselves to the case of a $\phi^3$
field-theory. Finally, we go beyond KB's work in casting our equations in a
form
convenient for computation by deriving a set of coupled equations for
two-fragment amplitudes in the $NN-\pi NN$ system.

This work \footnote{A summary of these results was presented at the
$14^{\rm th}$ International Conference on Few Body Problems in Physics
\cite{AP94B}.} proceeds as follows. In Section
\ref{sec-FTreview} Green's functions and amplitudes of the field theory are
defined while Section
\ref{sec-Taylorrev} provides a brief summary of the Taylor method of
classification of diagrams. In Section \ref{sec-NNpt1} the $NN$
amplitude $T_{NN}$ is discussed and is found to depend on the
fully-connected two-particle irreducible $NN \rightarrow \pi NN$
amplitude, ${F^{(2)}}^{\dagger}$. Section \ref{sec-F2adj} is therefore
devoted to deriving an equation for ${F^{(2)}}^{\dagger}$, a process
which involves eliminating the double-counting that occurs in the
equation found when the Taylor method is applied directly to
${F^{(2)}}^{\dagger}$. Section \ref{sec-NNdblectfix} then outlines the
removal of double-counting from the equation for the two-particle
irreducible amplitude $\tnni{2}$. Once in possession of
double-counting-free equations for the amplitudes involved we proceed in
Section \ref{sec-coupled} to derive a set of coupled equations for the
$NN-\pi NN$ system. Section \ref{sec-antisymm} then explains how
these equations may be anti-symmetrized in order to obtain the physical
amplitudes. Finally, Section \ref{sec-inpspec} gives suggestions for the
specification of the input amplitudes and explains how physics beyond the
one-pion sector can be implicitly included in the framework developed in
Sections \ref{sec-FTreview}--\ref{sec-antisymm}.

The work concludes with four appendices containing details
of arguments to do with the dressing of the propagators, the amplitudes
which are the input to the equations, and details of the double-counting
removal performed in Section~\ref{sec-NNdblectfix}.

\section  {The field theory of nucleons and pions}

\label {sec-FTreview}

In this section we define the field theory of nucleons and pions used
in this work by discussing the Lagrangian to be used and defining the
Green's functions and amplitudes of the theory.

\subsection {The Lagrangian}

The Taylor method is essentially a topological classification of the
diagrams in a perturbation expansion. Therefore, it depends only on
the topology of the diagrams which can be generated, and not on the
details of the particular Lagrangian under consideration. Consequently,
in this work we do not specify the precise form of the Lagrangian to be
used. We merely assume that
the Lagrangian density describes a system of nucleons and pions and
so takes the form:
\be
{\cal L}={\cal L_D} + {\cal L}_\phi +{\cal L}_{int},
\label {eq:Lag}
\ee
where:
\begin{eqnarray}  {\cal L}_D(x)=\overline{\psi}(x)
(i \gamma^\mu \partial_\mu - m) \psi&(&x),\\ {\cal
L}_\phi(x)=\frac{1}{2} (\partial_\mu \vec {\phi}(x)
\cdot \partial^\mu \vec{\phi}(x)  -  m_\pi^2
\vec{\phi}&(&x) \cdot \vec{\phi}(x)),
\end{eqnarray}
where $\psi$ is the nucleon field and $\vec{\phi}$ the pion field, which
is an isovector.

At this stage we do not actually need to assume {\em anything} about
the structure of ${\cal L}_{int}$ in order to apply the Taylor method.
However, for concreteness the reader may wish to assume that ${\cal
L}_{int}$ is a sum of a bare $NN \pi$ vertex, proportional to
$\bar{\psi} \phi \psi$ and a contact term, proportional to $\bar{\psi}
\phi \phi \psi$. As will  be seen below, our derivation never demands
that we say more than this. Bare form factors for these two kinds of
vertices could be included in the Lagrangian if we wished. Such form
factors would be a natural place to include information from QCD about
the structure of the nucleons and pions in our field theory. However,
for the present we leave these details to one side.

Note also that the Lagrangian independence of the Taylor method makes
it relatively easy to return at a later stage and add other mesons and
baryons to the field theory. The addition of extra particles to the
theory merely complicates the resulting equations without changing the
spirit of the derivation. However, as a first step in this
paper we have derived the results for a field theory in which the only
quanta are nucleons and pions.

\subsection {Green's functions and amplitudes}

The coordinate space Green's function for the transition
from a $j$-nucleon, $m-j$-pion state to a $j'$-nucleon,
$n-j'$-pion state is defined by:
\bea
G^{(j',j)}_{n
\leftarrow m}
&(&x_1',\ldots,x_{j'}',x_{j'+1}',\ldots,x_{n}';x_1,\ldots,x_{j},x_{j+1},
\ldots,x_m)=\nn\\
\langle 0|T&(&\psi(x_1') \ldots \psi(x_{j'}')
\phi (x_{j'+1}') \ldots \phi(x_n') \bpsi(x_1) \ldots
\bpsi(x_{j}) \phi(x_{j+1}) \ldots \phi(x_m))|0 \rangle,
\eea
where $|0 \rangle$ is the vacuum state, $T$ is the
time-ordering operator, $\psi(\bpsi)$ is the nucleon
annihilation (creation) operator, and $\phi$ is the pion
operator, whose isospin index has been suppressed.

This Green's function may then be Fourier transformed in
order to obtain the momentum-space Green's function:
\be
G^{(j',j)}_{n \leftarrow m}
(p_1',\ldots,p_{j'}',p_{j'+1}',\ldots,p_{n}';p_1,\ldots,p_{j},p_{j+1},\ldots,
p_m).
\ee
{}From this momentum-space Green's function the
amplitude:
\be
A^{(j',j)}_{n \leftarrow m}
(p_1',\ldots,p_{j'}',p_{j'+1}',\ldots,p_{n}';p_1,\ldots,p_{j},p_{j+1},\ldots,
p_m)
\ee
is obtained via the LSZ reduction \cite{LSZ55}
formula:
\bea
A^{(j',j)}_{n \leftarrow m}
(p_1',\ldots,p_{n}';p_1,\ldots,p_m)=
 Z_N^{-\frac{j'}{2}}
Z_\pi^{-\frac{n-j'}{2}} d_N^{-1}(p_1') \ldots
d_N^{-1}(p_{j'}') d_\pi^{-1}(p_{j'+1}') \ldots
d_\pi^{-1}(p_{n}')\nn\\
G^{(j',j)}_{n \leftarrow m}
(p_1',\ldots,p_n';p_1,\ldots,p_m)
d_N^{-1}(p_1) \ldots
d_N^{-1}(p_{j}) d_\pi^{-1}(p_{j+1}) \ldots d_\pi^{-1}(p_{m})
Z_N^{-\frac{j}{2}} Z_\pi^{-\frac{m-j}{2}}
\label {eq:LSZ}
\eea
where $d_N$ and $d_\pi$ are the free
single-nucleon and single-pion propagator in momentum
space, given by the formulae:
\bea
d_N(p)&=&\frac{i}{\not \! {p} - m_N},\\
d_\pi(k)&=&\frac{i}{k^2-m_\pi^2}.
\eea
Note that we are using fully dressed propagators in order to
do the reduction here and so $Z_N$ and $Z_\pi$ are the
wave function renormalizations for the nucleon and pion
respectively, while $m_N$ and $m_\pi$ are the dressed nucleon
and pion masses. The Green's function
$G^{(j',j)}_{n \leftarrow m}$, and therefore the amplitude $A^{(j',j)}_{n
\leftarrow m}$, may be expressed as the sum of a perturbation series of
Feynman diagrams.

Note that although all the formulae in this section
are written for identical particles, in this paper we (at
first) consider only the amplitudes for distinguishable
particles. The physically correct indistinguishable-particle
amplitudes may then be obtained from
these distinguishable-particle amplitudes by the usual
symmetrization and anti-symmetrization processes. This
procedure is fully justified and implemented in Section
\ref{sec-antisymm}.

\section {The Taylor method of
classification-of-diagrams}

\label {sec-Taylorrev}

\subsection {Taylor's original method}

Taylor's method of
classification-of-diagrams then provides a scheme for classifying all of
the diagrams which contribute to any amplitude $A^{(j',j)}_{n \leftarrow
m}$ according to their topology. In this paper we concentrate
on the topological structure of the diagrams in the
$s$-channel \footnote{We use the notation $s-$,$t-$ and $u$-channel
throughout
this paper, intending it in the sense defined by Mandelstam
\cite{Ma58,Ma59}.}, although it is possible to apply the Taylor procedure in
other channels instead of, or as well as, applying it in this channel. Note
that because of this focus on the $s$-channel structure,
unless otherwise specified, the irreducibilities given for
amplitudes and diagrams are $s$-channel irreducibilities. Note
also that from now on we suppress
the indication of the number of nucleons and pions present
in the initial and final states, in order to simplify the
notation.

Having noted these points the Taylor scheme then requires the
following definitions:
\begin{defn}[r-cut] An r-cut is an arc which separates
initial from final states and intersects exactly
$r$-lines, at least one of which must be an internal
line. If all of the r lines cut are
internal lines then the cut is called an {\em internal r-cut}.
\end{defn}

Note that in writing this definition we assume that all perturbation
diagrams are represented in a two-dimensional plane. We do allow the
lines in any diagram to ``jump over" one another: two lines  do not have
to meet at an interaction vertex whenever they intersect. By contrast, a
cut is defined to intersect all the lines it meets: it may not jump over
any of them. (Other definitions of an
$r$-cut, which do not assume that the diagrams are represented in the
plane, may be composed but it is the above definition which Taylor
himself used.)

\begin{defn}[r-particle irreducibility] A diagram is
called $r$-particle irreducible if, for all integers $0
\leq k \leq r$, no $k$-cut may be made on it. An
amplitude is called $r$-particle irreducible if all
diagrams contributing to it are $r$-particle irreducible.
\end{defn}
The $r$-cuts which may be made on a particular diagram are now
divided as follows: if an $r$-cut is not internal it is called
``initial" if it intersects at least one initial-state but  no
final-state line; ``final" if it intersects at least one final-state,
but no initial-state line, and ``mixed" if it intersects both initial
and final-state lines.

Once these definitions are made any diagram
contributing to the connected $s$-channel
$(r-1)$-particle irreducible ($r-1$PI) $m \rightarrow n$
amplitude, $A^{(r-1)}_{n \leftarrow m}$, may be placed in
one of the five classes $C_1$ to $C_5$. The class a particular graph
belongs to is determined by the $r$-cuts which may be made on it, as
follows:

\begin{list}%
{$C_{\arabic{class}}$:}{\usecounter{class}
\setlength{\rightmargin}{\leftmargin}}
\item No $r$-cut
may be made on the diagram, i.e. it is $r$-particle
irreducible;
\item At least one internal $r$-cut may be made
on the diagram and no mixed or final $r$-cut may be made.
\item At least one initial $r$-cut is possible on the diagram, but all
other types of cut are not possible;
\item At least one mixed $r$-cut is possible, but a final $r$-cut is
impossible;
\item At least one final $r$-cut may be made.
\end{list}
This classification is motivated by where the ``latest" $r$-cut may be
made on a particular graph. If it is an internal cut then the graph
must be placed in $C_2$, but if it is an external cut then the graph is
placed in one of $C_3$--$C_5$ depending on what sort of cut that
``latest" $r$-cut may be.

Because each graph contributing to $A$ belongs to one
and only one of these classes it follows that $A$ may be
expressed as the sum of the five expressions found by
summing the individual classes separately.

The sum of class $C_1$ is clearly the connected $s$-channel
$r$-particle irreducible $m \rightarrow n$ amplitude, $A^{(r)}_{n
\leftarrow m}$.
Each of the classes $C_2$ to $C_5$ may be summed using
the following result, which is known as the last internal
cut lemma (LICL):
\begin{lemma}[Last Internal Cut]
Any
$(r-1)$-particle irreducible diagram which admits an
internal $r$-cut has a unique internal $r$-cut which is
nearest to the final state.
\end{lemma}
For a proof of this result see paper I\@. Explicit results for
the sum of classes $C_3$--$C_5$ are also given there. Note also
the following results, more fully explained in paper I:
\begin{enumerate}
\item The Last Internal Cut Lemma (LICL) may not be applied directly
when constructing the sum of classes $C_3$--$C_5$;

\item In order to correctly sum classes $C_3$, $C_4$ and $C_5$ one
not only needs to restrict the $s$-channel cut-structure of the
amplitudes in the sums of these classes, but also to place
constraints upon the cut structure of these amplitudes in
channels other than the $s$-channel. (This is a point whose full
implications were apparently not realized by Taylor himself.)
However, many of these additional restrictions prove to be
trivially satisfied by the amplitudes under consideration, due
to the stringent nature of the $s$-channel irreducibilities
enforced by the LICL.
\end {enumerate}

\subsection {The double-counting problem and its solution}

\label {sec-dcsolve}

As was also explained in paper I, the adoption of Taylor's
procedure for summing the classes $C_3$ to $C_5$ leads to
the overcounting of certain diagrams, as follows.

In the case where $n \leq r$ there are complications with the
use of the last internal cut lemma in classes $C_4$ and
$C_5$. If $m<r$ similar difficulties arise in classes
$C_3$ and $C_4$. Taylor's solution to these difficulties
is to split any class $C$ in which such a problem occurs
into sub-classes $S$. These sub-classes are defined in such
a way that the last internal cut lemma may be applied to
the diagrams within them without complications, and
consequently their sum may be constructed correctly. But,
Taylor attempts to then construct the sum of $C$ by summing
over all possible sub-classes $S$. This is correct,
provided that the sub-classes  are defined so as to be
completely disjoint. However, with Taylor's definition of
sub-classes of $C$ the sub-classes are {\em not} disjoint.
Certain diagrams belong to more than one sub-class and so are
double-counted. In paper I the details of this problem
were discussed and specific examples given. It was then
shown how to avoid the double-counting problem by using the following
procedure to correctly construct the sum of any class $C$ in which
double-counting is a possibility.

The procedure involves considering each sub-class $S$ of
$C$ in turn. The sub-classes are added to a ``running tally"
of $C$, one by one. However, a particular sub-class $S$
may only be added to the running tally after all of the
diagrams within $S$ which have already been included in
the ``running tally" have been eliminated from $S$.

We express this mathematically by saying that if $C$ is the
union of $n$ sub-classes, i.e. $C=\cup_{i=1}^n S_i$, then the
correct sum of $C$ is:
\be
\sum_{i=1}^n S_i - \sum_{{\scriptstyle \begin{array}{c}
				      i,j=1\\
          i<j
          \end{array}}}^n S_i \cap S_j + \sum_{{\scriptstyle
\begin{array}{c}
				      i,j,k=1\\
          i<j<k
          \end{array}}}^n S_i \cap S_j \cap S_k - \ldots.
\label{eq:dcform}
\ee

The difficult part of the application of this formula is the
discovery of which diagrams are in the intersections $S_i \cap
S_j$, $S_i \cap S_j \cap S_k$, etc.\@. Because of the way the sub-classes
are defined any diagram which is in a set $S_i \cap S_j$ must admit two
possible ``unique" latest $r$-cuts, each cutting a different set of
external lines. Consequently, the diagrammatic content of the
set $S_i \cap S_j$ may be found by examining the full
$s$-channel cut-structure of the amplitude that is the sum of
$S_i$ and so determining which portions of that amplitude have a
cut-structure which means that they also belong to sub-class
$S_j$. In covariant perturbation theory, or, for that matter, in
any time-dependent perturbation theory, the result of this
examination will depend, not only on the $s$-channel cut-structure
of the sub-amplitudes making up the amplitude representing
$S_i$, but also on their cut-structure in other channels.
Therefore, once the double-counting is eliminated by doing the
subtraction recommended in Eq.~(\ref{eq:dcform}) certain
constraints are effectively placed on the cut-structure of these
sub-amplitudes in channels other than the $s$-channel. Thus, in
spite of the fact that we consider only the $s$-channel
cut-structure of the full amplitude, the elimination of
double-counting effectively places constraints on the
cut-structure of the sub-amplitudes in channels other than the
$s$-channel.

This modification to the Taylor method
provides a way of deriving covariant four-dimensional scattering
equations which are free of double-counting for any
system in which it is possible to obtain a graphical perturbation
expansion for the relevant amplitudes. The only {\em
caveat} is that the particles involved must be fully
dressed before the Taylor method is applied. In Appendix \ref{ap-dress}
we explain how this is accomplished for a system of nucleons and pions.
Also, in Appendix \ref{ap-input} the Taylor method is used to derive
equations for the amplitudes which are input to the $NN-\pi NN$
equations: the $\pi N$ amplitude and the $\pi N N$ vertex. In the next
section we demonstrate how the Taylor method as outlined here would be
applied to the $NN$ t-matrix, $T_{NN}$.

\section {The $N-N$ amplitude, part 1}

\label {sec-NNpt1}

Now consider the $N-N$ amplitude, $T_{NN}$. In this work we
are interested in particle-particle scattering, and so we may
consider the particles in the initial and final states to
both be nucleons, not anti-nucleons. Consequently, the baryon
number in any intermediate state must be two, and so the
amplitude $T_{NN}$ can only have an $s$-channel one-particle
irreducible part $\tnni{1}$. Furthermore, since, by assumption,
both nucleons are fully-dressed, this amplitude must be
connected. Taylor's method may therefore be applied to this
$T_{NN}^{(1)}$ in order to derive an integral equation for $NN$
scattering.

So consider any diagram contributing to $\tnni{1}$. Any cuts which
could be made to place this diagram in $C_3$--$C_5$ would merely
expose dressing contributions on the external nucleon legs. Since
all particles are fully dressed such dressing contributions must be zero,
and hence classes $C_3$--$C_5$ are empty. Meanwhile, class $C_1$ sums to
the connected two-particle irreducible $NN$ amplitude $T_{NN}^{(2)}$, and
class $C_2$ yields, using the LICL, the product $\tnni{2} d_1 d_2
\tnni{1}$ where $d_1$ and $d_2$ are the
fully-dressed free propagators for nucleons $1$ and $2$. Hence, Taylor's
method gives:
\be
T_{NN}^{(1)}=T_{NN}^{(2)} + \tnni{2} d_1 d_2 \tnni{1}.
\label {eq:BSNN1}
\ee

Eq.~(\ref{eq:BSNN1}) is, of course, a Bethe-Salpeter
equation for $\tnni{1}$, with $\tnni{2}$ the kernel of the equation,
containing the diagrams to be iterated in order to produce the full
amplitude. However, above the pion-production threshold
$\tnni{2}$ contains part of the effects of inelasticity. (The
fully-dressed nucleon propagators also contain some inelasticity
effects.) Consequently, if the $NN$ sector is to be modelled correctly
it is necessary that we examine the $NN \pi$ cut structure of $\tnni{2}$.
However, if this three-particle cut structure of $\tnni{2}$ is to be
examined via the Taylor method we have $m=n=2$ and $r=3$, so the
conditions for double-counting in classes $C_3$, $C_4$ and $C_5$ are
satisfied. Therefore, we must proceed with caution.

Applying Taylor's method directly to $\tnni{2}$ means that we
sum each class individually, as follows:

\begin{list}%
{$C_{\arabic{class}}$:}
{\usecounter{class}\setlength{\rightmargin}{\leftmargin}}
\item The sum of class $C_1$ is clearly the connected 3PI $2
\rightarrow 2$ amplitude, $\tnni{3}$.

\item Using the last-internal cut lemma we may show that
the sum of class $C_2$ is:
\be
F^{(3)} d_1 d_2 d_\pi
{F^{(2)}}^{\dagger},
\label {eq:T2C2}
\ee where $F^{(r)}$ is the connected
$r$-particle irreducible $NN \pi \rightarrow NN$
amplitude, and $\bfai{r}$ is the connected $r$-particle
irreducible $NN \rightarrow NN \pi$ amplitude. Now, $C_2$ may not
contain any diagrams admitting a three-cut which involves a final-state
line. It might be thought that in order to enforce this condition in
Eq.~(\ref{eq:T2C2}) certain restrictions would need to be placed on
$F^{(3)}$. However, when these extra conditions are examined
closely it is found that they are all satisfied trivially, due to
the $s$-channel three-particle irreducibility of $F^{(3)}$. (This
is, in fact, a special case of a more general result, that in
$C_2$ such extra conditions are always unnecessary, because the
cuts which may intersect a final-state line and so place diagrams in
$C_4$ or $C_5$ are precluded by the $s$-channel cut-structure imposed on
the amplitudes involved.)

\item As explained in paper I, the sum of class $C_3$
must be constructed with care. This is because the LICL
applies only to internal cuts, and so cannot be used
directly on diagrams in $C_3$, since, by definition, diagrams
in $C_3$ may have cuts which intersect both internal and
external lines.

In order to overcome this problem Taylor constructs two
sub-classes of $C_3$, one of which contains those
diagrams in which the ``last" cut cuts the line $N1$, and
the other of which contains those diagrams in which the
``last" cut cuts the line $N2$. (Here the sense in which
the cut we are talking about is the ``last" cut was
defined in paper I\@.) Taylor calls the first sub-class
$C_3^{\{N1\}}$ and the second $C_3^{\{N2\}}$. Using the
LICL he constructs their sums:
\bea
C_3^{\{N1\}}=F^{(3)} d_2 d_\pi \fai{1}(1), \label
{eq:C3N1}\\
C_3^{\{N2\}}=F^{(3)} d_1 d_\pi \fai{1}(2), \label
{eq:C3N2}
\eea
where $\fai{1}(i)$ is the 1PI $\pi NN$ vertex
corresponding to the emission of a pion by nucleon
$\bar{i}$ while nucleon $i$ is  a spectator. I.e.,
$\bar{i}$ is defined by:
\begin{equation}
\bar{i}=\left \{ \begin{array}{ll}
 2 & \mbox { if $i=1$}\\ 1 & \mbox { if $i=2$.}
\end{array}  \right.
\end{equation}
(Note that once again it might be expected that conditions must be
placed on $F^{(3)}$ in order to ensure that no diagram which
actually belongs in $C_4$ or $C_5$ is included in $C_3$. And,
indeed, as we shall see below, in some cases the amplitudes used
in $C_3$ do have to be restricted. But, in the case under discussion
here the $s$-channel 3PI of $F^{(3)}$ guarantees that only diagrams which
belong in $C_3$ are included in the sums (\ref{eq:C3N1}) and
(\ref{eq:C3N2}).)

Taylor then claims that the sum of $C_3$ is merely the
sum of the two sub-class sums (\ref{eq:C3N1}) and
(\ref{eq:C3N2}). But this is false, since certain
diagrams will have two distinct possible ``latest" cuts,
one cutting $N1$ and one cutting $N2$. (See, e.g. Figure
\ref{fig-C3dc}.) These diagrams will belong to both
sub-classes, and so will be double-counted in such a
summation. In order to correct this double-counting we apply
Eq.~(\ref{eq:dcform}) so as to obtain the corrected sum of
class $C_3$:
\be
C_3=\sum_{i=1}^2 F^{(3)} d_{\bar{i}} d_\pi \fai{1}(i) - D_3,
\ee
where:
\be
D_3=C_3^{\{N1\}} \cap C_3^{\{N2\}},
\ee
has been subtracted in order to remove the double-counting.

\item  Adopting a similar approach to that used to sum
$C_3$, we follow Taylor, and divide $C_4$ into two
sub-classes: $C_4^{\{N2'\}\{N1\}}$, which contains all those
diagrams in which the ``latest" cut cuts lines $N2'$ and
$N1$, and $C_4^{\{N1'\}\{N2\}}$, which contains all
those diagrams in which the ``latest" cut cuts lines $N1'$
and $N2$. The last internal cut lemma may then be employed
in order to write:
\bea
C_4^{\{N2'\}\{N1\}}&=&f^{(2)}(2) d_\pi
\tfai{1}(1), \label{eq:c421}\\
C_4^{\{N1'\}\{N2\}}&=&f^{(2)}(1)
d_\pi \tfai{1}(2).   \label{eq:c412}
\eea
Note that in order to stop diagrams which should be in $C_5$
straying into $C_4$ we have had to restrict the amplitude
$\fai{1}$ used in (\ref{eq:c421}) and (\ref{eq:c412}) to only
contain diagrams which are 2PI in the $N' \leftarrow N + \pi'$-channel.
We have denoted the resulting amplitude by $\tfai{1}$. If the
restriction on $\fai{1}$ is not enforced then the three-cuts
depicted in Figure \ref{fig-C4cuts} may be made on certain
diagrams in $C_4$, indicating that these diagrams actually belong
in $C_5$, not $C_4$

If we now apply Eq.~(\ref{eq:dcform}) we find that the correct
sum of $C_4$ is:
\be
C_4=\sum_{i,j=1}^2 f^{(2)}(j) d_\pi
\tfai{1}(i)
\bar{\delta}_{ij}-D_4,
\ee
with:
\be
D_4=C_4^{\{N2'\}\{N1\}} \cap C_4^{\{N1'\}\{N2\}}.
\label{eq:D4}
\ee

\item Once more this class is split into sub-classes.
This time the sub-classes are $C_5^{\{N1'\}}$ and
$C_5^{\{N2'\}}$, which contain, respectively, all
diagrams in which the ``latest" cut cuts lines $N1'$ and
$N2'$. The LICL gives:
\be C_5^{\{Nj'\}}=\fir{2}(j) d_{\bar{j}}
d_\pi {F^{(2)}}^{\dagger},
\ee
for $j=1,2$. Consequently we have:
\be
C_5=\sum_{j=1}^2 \fir{2}(j) d_\pi d_{\bar{j}}
{F^{(2)}}^{\dagger}-D_5,
\ee
with:
\be
D_5=C_5^{\{N1'\}} \cap C_5^{\{N2'\}}.
\label {eq:D5}
\ee
\end {list}

If we combine the above results for the sums of the individual
classes we find:
\bea
T_{NN}^{(2)} &=& T_{NN}^{(3)} + F^{(3)} \gtb ({F^{(2)}}^{\dagger} +
\sum_i \fai{1}(i) d_i^{-1}) - D_3\nn\\
&& + \sum_{j} f^{(2)}(j)  d_{\bar{j}} d_\pi (\sum_i \bd{i}{j} \tfai{1}(i)
d_i^{-1} + {F^{(2)}}^{\dagger}) - D_4 - D_5
\label{eq:T2}
\eea
where here, and throughout the rest of the paper, the sums over $i$ and $j$
are understood to run over $i,j=1,2$. This equation, but without the
restriction
on $\fai{1}$ in class $C_4$ and the subtractions for
double-counting, was also derived by Avishai and Mizutani
\cite{AM83} and Afnan and Blankleider
\cite{AB80,AB85}. Below we shall
calculate the precise values of the subtraction terms $D_3$,
$D_4$ and $D_5$. But, in order to achieve this and so derive an
equation for $\tnni{2}$ we will need to know the structure of the
amplitudes in Eq.~(\ref{eq:T2}).

Firstly, consider the amplitudes $T_{NN}^{(3)}$ and $F^{(3)}$. Explicit
examination of the structure of these amplitudes would involve exposing
$NN \pi \pi$ intermediate states. Since our primary concern here is the
derivation of equations which treat the $NN \pi$ intermediate states of
the theory correctly, we argue that these amplitudes may be safely
neglected or approximated. The equations resulting from this
approximation will then satisfy $NN$ and $NN \pi$ unitarity, but not $NN
\pi \pi$ unitarity. As a first approximation we set $F^{(3)}$ and
$T_{NN}^{(3)}$ to zero, in order to make our equations as simple as
possible. This approximation is, in fact, exact in the case of an $N
\pi$ interaction Lagrangian containing only a $\pi NN$ vertex.

The use of this approximation means that the sum of both sub-classes of
$C_3$ is zero, and so $D_3=0$. Therefore, given this
assumption, $T_{NN}^{(2)}$ obeys:  \be
T_{NN}^{(2)}=\sum_j \fir{2}(j)
d_{\bar {j}} d_\pi {F^{(2)}}^{\dagger} + \sum_{ij}
\bar{\delta}_{ij} \fir{2}(j) d_\pi \tfai{1}(i)-D_4-D_5,
\label{eq:T22}
\ee
with $D_4$ and $D_5$ given by equations (\ref{eq:D4}) and
(\ref{eq:D5}) respectively. The amplitude $\fir{2}$ which
appears in (\ref{eq:T22}) can be extracted from the model
Lagrangian, as demonstrated in Appendix \ref{ap-input}. As far as the
amplitude $\tfai{1}$ is concerned, for the moment we merely observe that
it must contain some subset of the diagrams which are summed by
solving Eq.~(\ref{eq:vertex}) for $\fir{1}$. Therefore the only
amplitude whose structure remains to be investigated is
${F^{(2)}}^{\dagger}$.

\section {The two-particle irreducible, $NN \rightarrow
NN \pi$ amplitude, ${F^{(2)}}^{\dagger}$}

\label {sec-F2adj}

Our examination of the two-particle irreducible $NN$
amplitude $\tnni{2}$ revealed that the equation it obeyed
involved the connected 2PI $NN \rightarrow NN \pi$ amplitude
$\bfai{2}$. In this section we examine this amplitude
using the original Taylor method, and eliminate the
double-counting inherent in that method, using the
modification to the Taylor method described in Section
\ref{sec-dcsolve} above.

In this case the sum of the Taylor classes and sub-classes
is as follows:
\bea
C_1&=&\bfai{3}\\
C_2&=&M^{(3)} \gtb \bfai{2}\\
C_3^{\{N1\}}&=&M_{1}^{(3)} d_2 d_\pi \fai{1}(1)\\
C_3^{\{N2\}}&=&M_{2}^{(3)} d_1 d_\pi \fai{1}(2)\\
C_4^{\{N1'\}\{N2\}}&=&\tpni{2}(1) d_\pi \fai{1}(2) \label{eq:C412}\\
C_4^{\{N2'\}\{N1\}}&=&\tpni{2}(2) d_\pi \fai{1}(1) \label{eq:C421}\\
C_4^{\{\pi'\}\{N1\}}&=&\tnni{2} d_\pi \fai{1}(1)
\label{eq:C4pi1}\\
C_4^{\{\pi'\}\{N2\}}&=&\tnni{2} d_\pi
\fai{1}(2) \label{eq:C4pi2}\\
C_5^{\{N1'\}}&=&\tpni{2}(1) d_2
d_\pi \bfai{2}\\ C_5^{\{N2'\}}&=&\tpni{2}(2) d_1 d_\pi
\bfai{2}\\
C_5^{\{\pi'\}}&=&\tnni{2} d_1 d_2 \bfai{2},
\eea
where we
have labeled the incoming nucleons $N1$ and $N2$, the outgoing
nucleons $N1'$ and $N2'$, the outgoing pion $\pi'$, and
$\tpni{2}(i)$ is the 2PI $\pi-N$ t-matrix with nucleon
$i$ as a spectator.
Here, $M^{(3)}$, $M_1^{(3)}$ and $M_2^{(3)}$ are all
connected 3PI $NN \pi \rightarrow NN \pi$ amplitudes. However, constraints
beyond simple $s$-channel 3PI have been placed on the amplitudes $M_1^{(3)}$
and
$M_2^{(3)}$. These constraints mean that $M_1^{(3)}$ and $M_2^{(3)}$ are
different and neither is equal to the sum of all $s$-channel 3PI $NN \pi
\rightarrow NN \pi$ diagrams, $M^{(3)}$. It is found that in order to stop
diagrams actually belonging to $C_4$ being mistakenly included in $C_3$ we must
define $M^{(3)}_i$ to be 1PI in both the channels:
$$Ni' + Ni \leftarrow  N\bar{i}' + N\bar{i} + \pi' + \pi$$
and:
$$\pi' + Ni \leftarrow N\bar{i}' + N\bar{i} + Ni' + \pi.$$
It might appear that conditions beyond $s$-channel 3PI would also have to be
placed on the $M^{(3)}$ appearing in $C_2$, in order to prevent diagrams from
$C_4$ and $C_5$ being included there too. However, in the case under
consideration in this section the possibility of any cut placing a diagram
which is summed in $C_2$ above in $C_4$ or $C_5$ is precluded by the
$s$-channel
3PI of $M^{(3)}$. Therefore, as was the case with the amplitude $F^{(3)}$ used
 in
$C_2$ and $C_3$ in the previous section, no constraints beyond $s$-channel
3PI are necessary.

Note that once these restrictions are imposed on $M_i^{(3)}$ the problem
discussed by Kvinikhidze and Blankleider \cite{BK94B}, where diagrams hidden in
$M^{(3)} d_{\bar{i}} d_\pi \fai{1}(i)$ actually represent processes
included in $C_4$, does not occur, since the diagrams in $C_4$ which
give rise to these difficulties are specifically excluded from
$C_3^{\{N1\}}$ and $C_3^{\{N2\}}$ by the above irreducibility
conditions. Therefore we may safely conclude that all diagrams in $C_2$ and
$C_3$
will {\em only} represent processes involving some piece of the connected
$s$-channel 3PI $NN \pi \rightarrow NN\pi$ amplitude. In this paper any such
process is said to involve a ``pure three-body force". In the calculation here
we assume that to a first approximation these pure three-body forces are
negligible, by an appeal to the same argument which resulted in us neglecting
the amplitudes $F^{(3)}$ and $T^{(3)}$ above. Consequently, we assume that the
sum of classes $C_2$ and $C_3$ is zero.

The inclusion of a three-body force, $M^{(3)}$ in our calculation,
would result in considerable changes to the
ensuing argument, since at numerous stages during
our discussion below we have appealed to the absence of a three-body
force as a reason for eliminating certain double-counting subtraction
terms. Nevertheless, integral equations for the amplitudes involved may
still be derived, even if $M^{(3)} \neq 0$. In that case, however, it is
considerably harder to see how such equations may be manipulated into a
form for practical calculation.

Once these observations have been made it is a simple
matter to sum the above expressions for the sub-classes in order
to obtain the following equation for $\bfai{2}$:
\be
\bfai{2}=\sum_{\alpha} t^{(2)}(\alpha) d_\alpha^{-1} \gtb
(\sum_i \bar{\delta_{i \alpha}} \fai{1}(i) d_i^{-1} + \bfai{2})
- D_4
\ee
with the sum over $\alpha$ running over $\alpha=1,2,3$ here and throughout
the rest of this work; $t^{(2)}(\alpha)$ being defined by:
\begin{equation}
t^{(2)}(\alpha)=\left \{ \begin{array}{ll}
 \tpni{2}(i) & \mbox { if $\alpha=i=1,2$}\\
 \tnni{2}    & \mbox { if $\alpha=3$.}
\end{array}  \right.
\label{eq:def1}
\end{equation}
and:
\bea
D_4&=&\cfo{\pi'}{N1} \cap \cfo{\pi'}{N2} + \cfo{N2'}{N1} \cap
\cfo{\pi'}{N2} + \cfo{N1'}{N2} \cap \cfo{\pi'}{N1}\nn\\
&+& \cfo{N2'}{N1} \cap \cfo{\pi'}{N1} + \cfo{N1'}{N2} \cap
\cfo{\pi'}{N2} + \cfo{N1'}{N2} \cap \cfo{N2'}{N1},
\eea
included in order to eliminate the double-counting in  $C_4$, as
is prescribed in Eq.~(\ref{eq:dcform}).  Note three things about
this result:
\begin{enumerate}
\item The $D_4$ here is obviously different to the $D_4$ defined
by Eq.~(\ref{eq:D4}) in the previous section, just as the sums of the
Taylor classes in this section are different since we are dealing with a
different amplitude here.

\item The maximum number of final-state lines cut by any three-cut is
$s_f=1$. Hence the condition $n \geq 2s_f$ holds and this condition is
sufficient to guarantee the absence of double-counting in $C_5$. (See
paper I for details.)

\item All three-sub-class intersections are zero in this particular case,
thus $D_4$ is given by the sum of the two-sub-class intersections
listed here.
\end {enumerate}

We now calculate the six two-set intersections involved in $D_4$. In
order to do this we note that any diagram which is in both
$\cfo{A'}{B}$ and $\cfo{E'}{F}$ must admit two ``latest" three-cuts,
the first of which cuts the three lines $A',B$ and one other, and
the second of which cuts the lines $E',F$ and one other.
Consequently, the intersection of $\cfo{A'}{B}$ and $\cfo{E'}{F}$
may be found by investigating which diagrams contributing to the
sum $\cfo{A'}{B}$ admit a three-cut involving $E'$, $F$ and one
other particle.

Looking at the sum of $\cfo{\pi'}{N1}$ we see that no three-cut
involving $\pi'$, $N2$ and one other particle is possible.
Therefore we conclude:
\be
\cfo{\pi'}{N1} \cap \cfo{\pi'}{N2}=0.
\ee
Similarly:
\be
\cfo{N2'}{N1} \cap \cfo{\pi'}{N1}=0,
\ee
and:
\be
\cfo{N1'}{N2} \cap \cfo{\pi'}{N2}=0.
\ee

Now consider $\cfo{N2'}{N1}$. Upon examining the diagrammatic sum
of $\cfo{N2'}{N1}$ we observe that there are a number of diagrams
contributing to the sum in which a three-cut involving $\pi'$ and
$N2$ may be made (see Fig.~\ref{fig-dcviau1pr}). Such a cut
may be made in the portion of $\cfo{N2'}{N1}$ which sums to:
\be
v_{\pi N}^X (2) d_\pi \fai{1}(1),
\ee
where $v_{\pi N}^X$ is the $u$-channel 1PR part of
$\tpni{2}$. In Appendix \ref{ap-input} it is shown that
the part of
$t_{\pi N}^{(1)}$ which is one-particle reducible in the
$s$-channel is the pole diagram, $\fai{1} d_N \fir{1}$.
By changing the argument of Appendix \ref{ap-input} to apply to
the $u$-channel instead of the
$s$-channel it may be shown that
$v_{\pi N}^X$ is merely the crossed term:
\be
v_{\pi N}^X=\fir{1} d_N \fai{1}.
\ee
Therefore,
\bea
\cfo{N2'}{N1} \cap \cfo{\pi'}{N2}&=&v^X(2) d_\pi \fai{1}(1)
\label{eq:dcx1}\\
&=& v_{OPE} d_1 \fai{1}(2),
\eea
Note that here $v_{OPE}$ is the full one-pion
exchange potential:
\be
v_{OPE}=\fir{1}(2) d_\pi \fai{1}(1).
\ee
This argument makes it explicit how the term (\ref{eq:dcx1}) comes
to be in two sub-classes. It is this inclusion of (\ref{eq:dcx1}) in
two different sub-classes which leads to the double-counting pointed out
by Kowalski et al. \cite{Ko79}.

Naturally exactly the same argument, but with the roles of $N1$
and $N2$ reversed, may be used to show that:
\bea
\cfo{N1'}{N2} \cap \cfo{\pi'}{N1}&=&v^X(1) d_\pi
\fai{1}(2) \label{eq:dcx2}\\
&=& v_{OPE} d_2 \fai{1}(1).
\eea
The identification in the last line may be made because the
one-pion exchange potential may be written as:
\be
v_{OPE}=\fir{1}(1) d_\pi \fai{1}(2),
\ee
since we are working in a time-dependent perturbation theory, and
so:
\be
\fai{1}=\fir{1}.
\label {eq:verteq}
\ee
Note that in order for Eq.~(\ref{eq:verteq}) to hold all
particles involved must be fully dressed, so that $\fir{1}$ is
one-particle irreducible in all channels.

Finally we attempt to construct the intersection of
$\cfo{N1'}{N2}$ and $\cfo{N2'}{N1}$. Any diagram which is in
$\cfo{N1'}{N2}$ and admits an $s$-channel three-cut involving
$N2'$, $N1$ and one other particle will also be in
$\cfo{N2'}{N1}$. As is shown in Fig.~\ref{fig-dcviat1pr} such a cut
is possible if the amplitude $\tpni{2}$ used in $\cfo{N1'}{N2}$ is
one-particle reducible in the $t$-channel. Therefore in models
including diagrams such as Fig.~\ref{fig-pipiia} the
intersection will be given by:
\be
\cfo{N1'}{N2} \cap \cfo{N2'}{N1}=v_{\pi N}^{\rho} (1) d_\pi
\fai{1}(2),
\ee
where $v_{\pi N}^{\rho}$ is the $t$-channel one-particle reducible
part of the $s$-channel 2PI $\pi$-N amplitude. However, in our
model pions do not interact with any particle other than nucleons,
and so the $\pi-N$ t-matrix is automatically one-particle
irreducible in the $t$-channel. Consequently, the cut shown in
Fig.~\ref{fig-dcviat1pr} is simply not possible, leading to:
\be
\cfo{N1'}{N2} \cap \cfo{N2'}{N1}=0.
\ee

Therefore we have:
\bea
D_4&=&\sum_{ij} \fir{1}(i) d_j \fai{1}(i) d_\pi \fai{1}(j) \bd{i}{j}
\label{eq:D4res1}\\
&=& \sum_{ij} v^X(i) d_\pi \fai{1}(j) \bd{i}{j}\\
&=& \sum_{ij} v_{OPE} d_j \fai{1}(i) \bd{i}{j}
\eea
As these rewritings show, when this result for $D_4$ is substituted into the
above integral equation for $\bfai{2}$ the subtracted diagrams may either be
used to modify the $\pi-N$ t-matrix or the $N-N$ t-matrix. As
will be seen below, which t-matrix we choose to modify here will
affect our later results for other amplitudes. In this work we choose the
former approach, however we stress that modifying the $N-N$ t-matrix is
equally legitimate. When these subtractions are made we find that:
\bea
\bfai{2}=\sum_{\alpha i} v^R(\alpha) d_\alpha^{-1} \bd{\alpha}{i}
d_{\bar{i}} d_\pi \fai{1}(i) + \sum_\alpha t^{(2)}(\alpha) d_\alpha^{-1} \gtb
\bfai{2}
\label{eq:F2adj}
\eea
with:
\begin{equation}
v^R(\alpha)=\left \{ \begin{array}{ll}
 \tpni{2}(i) - v^X(i) & \mbox { if $\alpha=i=1,2$}\\
 \tnni{2}    & \mbox { if $\alpha=3$.}
\end{array}  \right.
\label{eq:vR}
\end{equation}

Note that:
\begin{enumerate}
\item If we choose to modify the $NN$ t-matrix Eq.~(\ref{eq:F2adj})
still holds, but with $v^R(\alpha)$ given by:
\begin{equation}
v^R(\alpha)=\left \{ \begin{array}{ll}
 \tpni{2}(i) & \mbox { if $\alpha=i=1,2$}\\
 \tnni{2} - v_{OPE}   & \mbox { if $\alpha=3$.}
\end{array}  \right.
\label{eq:altdecomp}
\end{equation}

\item Provided we use amplitudes $\tnni{2}$ and
$\tpni{2}$ which themselves contain no double-counting the
integral equation for $\bfai{2}$, Eq.~(\ref{eq:F2adj}), is completely
correct and, in turn, contains no double-counting.

\item There
is a  bootstrap problem here. We set out to determine $T_{NN}^{(2)}$
and found it depended on $\bfai{2}$. Now we find that
$\bfai{2}$ depends on $T_{NN}^{(2)}$!
\end {enumerate}

However, leaving this bootstrap problem aside for the moment,
Eq.~(\ref{eq:F2adj}) may be iterated in order to sum the
multiple-scattering series for $\bfai{2}$. If this is done and the
definitions (\ref{eq:def1})--(\ref{eq:deflast}) used, we obtain:
\bea
\bfai{2}&=&\sum_{\beta i} \tit{1}(\beta) d_\beta^{-1} \gtb \bar
{\delta}_{\beta i} \fai{1}(i) d_i^{-1}\nn\\ &+& \sum_{\alpha \beta i}
t^{(1)}(\alpha) d_\alpha^{-1} \gtb U_{\alpha \beta}^{(2)} \gtb
\tit{1}(\beta) d_\beta^{-1} \gtb \bar{\delta}_{\beta i} \fai{1}(i)
d_i^{-1},
\label{eq:bfa2}
\eea
where the $U_{\alpha \beta}^{(2)}$s obey the four-dimensional covariant AGS
equations \cite{AGS67}:  \bea
U_{\alpha \beta}^{(2)}=\bar {\delta}_{\alpha \beta} d_1^{-1}
d_2^{-1} d_\pi^{-1} + \sum_{\gamma}
\bar{\delta}_{\alpha \gamma} t^{(1)}(\gamma)
d_\gamma^{-1} \gtb U_{\gamma \beta}^{(2)},
\label{eq:AGS1}\\
=\bar{\delta}_{\alpha \beta} d_1^{-1} d_2^{-1} d_\pi^{-1} +
\sum_{\gamma} U_{\alpha \gamma}^{(2)} \gtb
t^{(1)}(\gamma) d_\gamma^{-1} \bar {\delta}_{\gamma
\beta},
\label{eq:AGS2}
\eea
and:
\bea
\ttpni{1} \equiv (1+\tpni{1} d_N d_\pi) v^R_{\pi N}&=&v^R_{\pi N} +
\tpni{2} d_N d_\pi \ttpni{1}, \label {eq:tildet2}\\
\tilde{t}_{NN}^{(1)}&\equiv&\tnni{1} \label{eq:deflast},
\eea
where $T_{NN}^{(1)}$ is, in principle, the full $NN$ t-matrix,
which is related to $\tnni{2}$ by Eq.~(\ref{eq:BSNN1}).

Note that all two-body amplitudes in this series are the
full $s$-channel 1PI two-body amplitudes $t^{(1)}$, except that
the first
$\pi-N$ scattering after the pion emission involves a
t-matrix, $\ttpni{1}$, which has had the crossed term
removed from the potential. This removal of the crossed
term is the only way in which this result differs from
that obtained by Avishai and Mizutani and Afnan and
Blankleider. It is necessary because, as was first
pointed out by Kowalski et al. \cite{Ko79}, the inclusion
of the crossed diagram in the first $\pi-N$ amplitude after
pion emission leads to double-counting of terms such
as Eq.~(\ref{eq:dcx1}) and (\ref{eq:dcx2}) if the calculation is
done in a time-dependent perturbation theory. Because we have used
the modified Taylor method, which is specifically
designed {\em not} to lead to double-counting, the above
equation does not contain such double-counting, even though it
was derived in a time-dependent perturbation theory.

\section {The $N-N$ amplitude, part 2}

\label {sec-NNdblectfix}

Having obtained the correct expression for $\bfai{2}$ we
may now return to our calculation of $\tnni{2}$. Recall
that above we derived Eq.~(\ref{eq:T22}) for $\tnni{2}$,
with the first term generated by $C_5$ and the second by
$C_4$. However, recall also that both terms were expected to
contain double-counting. Having derived an expression for
$\bfai{2}$ which is itself free of double-counting we may now
determine the factors $D_4$ and $D_5$, which were introduced into
Eq.~(\ref{eq:T22}) in order to compensate for this
double-counting.

\subsection {Calculating $D_4$}

Firstly, note that:
\be
D_4=\cfo{N1'}{N2} \cap \cfo{N2'}{N1},
\ee
with $\cfo{N1'}{N2}$ and $\cfo{N2'}{N1}$ given by the
expressions (\ref{eq:c421}) and (\ref{eq:c412}).

Now observe that since all particles are fully dressed all vertices
are at least 1PI in all channels. Therefore in a time-dependent
perturbation theory:
\be
\tfai{1}=f^{(2)}.
\label{eq:tdpt1}
\ee
Consequently:
\be
f^{(2)}(2) d_\pi \tfai{1}(1)=f^{(2)}(1) d_\pi
\tfai{1}(2),
\ee
i.e.\@ one sub-class merely reproduces the sum of the other.
Topologically this happens because the two cuts $N1' \pi_I N2$
and $N2' \pi_I N1$ (here $I$ stands for intermediate line) may
be made on {\em any} diagram in $C_4$ and so all diagrams in $C_4$
belong to both sub-classes. Therefore:
\bea
D_4&=&\cfo{N1'}{N2}=f^{(2)}(2) d_\pi
\tfai{1}(1)=\cfo{N2}{N1'}=f^{(2)}(1) d_\pi \tfai{1}(2).
\label {eq:D4res}
\eea

\subsection {Calculating $D_5$}

\label{sec-calcD5}

In Section \ref{sec-NNpt1} we defined $D_5$ to be:
\be
D_5=\cfi{N1'} \cap \cfi{N2'}.
\ee
In order to discover which diagrams are actually in $D_5$ we
examine the sum of $C_5^{\{N1'\}}$:
\be
\cfi{N1'}=\fir{2}(1) d_2 d_\pi \bfai{2},
\label {eq:C51}
\ee
and note which diagrams in $\cfi{N1'}$ admit a three-cut
involving $N2'$ and so should be included in $D_5$. Naturally,
provided that we use an expression for $\bfai{2}$ which is
itself free of double-counting, such as Eq.~(\ref{eq:bfa2})
derived above, the sum of $\cfi{N1'}$ will itself be free of
double-counting. Consequently, such a procedure will correctly
identify all of the diagrams which must be included for
subtraction in $D_5$.

Now substituting in (\ref{eq:C51}) for $\bfai{2}$ from
Eq.~(\ref{eq:bfa2}) and iterating using the AGS equations gives:
\bea
&\cfi{N1'}&=\sum_{\alpha i} \fir{1}(1) \bd{1}{\alpha} d_2 d_\pi
\tit{1}(\alpha) d_\alpha^{-1} d_{\bar{i}} d_\pi
\bd{\alpha}{i} \fai{1}(i)  + \fir{2}(1) d_2 d_\pi \ttpni{1}(1)
d_\pi \fai{1}(2)\nn\\
&+& \sum_{\alpha \beta i} \fir{1}(1)
\bd{1}{\alpha} d_2 d_\pi t^{(1)}(\alpha) d_{\alpha}^{-1} \gtb
U_{\alpha \beta}^{(2)} \gtb \tit{1}(\beta) d_\beta^{-1}
d_{\bar{i}} d_\pi \bd{\beta}{i} \fai{1}(i).
\eea
We write:
\be
\cfi{N1'}=c_1+c_2+c_2+c_3+c_4+c_5+c_6,
\ee
where:
\bea
c_1&=&\fir{1}(1) d_\pi \ttpni{1}(2) d_2 d_\pi \fai{1}(1),
\label{eq:c1}\\ c_2&=&\fir{1}(1) d_2 d_\pi \tnni{1} d_2 \fai{1}(1),\\
c_3&=&\fir{1}(1) d_2 d_\pi \tnni{1} d_1
\fai{1}(2),\label{eq:c3}\\
c_4&=&\fir{2}(1) d_2 d_\pi
\ttpni{1}(1) d_\pi \fai{1}(2),\\
c_5&=&\sum_{\beta i} \fir{1}(1)
d_2 d_\pi \tpni{1}(2) d_1 d_\pi U_{2
\beta}^{(2)} \gtb \tit{1}(\beta) d_\beta^{-1} \bd{\beta}{i}
d_{\bar{i}} d_\pi \fai{1}(i),\\
c_6&=&\sum_{\beta i} \fir{1}(1) d_2 d_\pi \tnni{1} d_1 d_2 U_{3
\beta}^{(2)} \gtb \tit{1}(\beta) d_\beta^{-1} \bd{\beta}{i}
d_{\bar{i}} d_\pi \fai{1}(i). \label{eq:c6}
\eea
It is clear that:
\be
D_5=\sum_{a=1}^6 c_a \cap \cfi{N2'}.
\label{eq:sumint}
\ee
Therefore in order to determine the value of $D_5$ it is
necessary to examine the six terms $c_1$--$c_6$ individually,
probing carefully in order to determine which diagrams from each
term admit a three-cut involving $N2'$.

\subsubsection {$c_1$}

\label {sec-c1calc}

We wish to calculate:
\be
c_1 \cap \cfi{N2'}.
\ee
In order to do this we observe that if the diagram representing
$c_1$ is to admit an $s$-channel three-cut involving $N2'$ and two
internal lines then the amplitude $\ttpni{1}$ will have to be
two-particle reducible in the $N' \leftarrow \pi' + \pi + N$ channel
(see Fig.~\ref{fig-firstterm}). Therefore we investigate
$\ttpni{1}$ in order to discover what portion of it is
two-particle reducible in this channel. We begin by using
Eq.~(\ref{eq:tildet2}) to write:
\be
\ttpni{1}=v^R + v^X d_N d_\pi \ttpni{1} + v^R d_N d_\pi
\ttpni{1}.
\label{eq:ttexp}
\ee
The second term here is 2PR in the channel $N' \leftarrow \pi'
+ \pi + N$. Therefore,
substituting this term into Eq.~(\ref{eq:c1}) for $c_1$ it is found that
the diagram: \be
\fir{1}_{\pi'}(1) d_{\pi'} \fir{1}_{\pi}(2) d_1 \fsai{1}{\pi'}(2)
d_1 d_\pi \ttpni{1}(2) d_2 d_\pi \fsai{1}{\pi}(1)
\ee
is in $c_1 \cap \cfi{N2'}$.

Examining (\ref{eq:ttexp}) it is now seen that the only other way in
which $\ttpni{1}$ could be 2PR in the $N'  \leftarrow \pi' + \pi + N$-channel
is if $v^R$ is 2PR in the same channel. Therefore we now examine
the effect of placing the $N'  \leftarrow \pi' + \pi + N$-channel 2PR
part of $v^R$ in the expression for $c_1$.

\begin{claim}
The portions  of the expression:
$$ \fir{1}(1) d_\pi \tilde{v}^{(2)}(2) (1 + d_1 d_\pi \ttpni{1}(2)) d_2
d_\pi \fai{1}(1), $$
where $\tilde{v}^{(2)}$ is the $u$-channel 1PI, $s$-channel 2PI and $N'
\leftarrow \pi' + \pi + N$-channel 2PR $\pi-N$ interaction, which are
also in $C_5^{\{N2'\}}$, given the assumptions of this calculation, are:
\be
\fir{1}_{\pi_2}(1) d_{\pi_2} \fir{1}_{\pi_1}(2) d_1 t^{(1)}_{\pi_2 N}(2)
d_{\pi_1} d_1 \fsai{1}{\pi_1}(2) d_{\pi_2} d_2 \fsai{1}{\pi_2}(1).
\label {eq:D2}
\ee
and:
\bea
\fir{1}_{\pi_2}(1) d_{\pi_2} d_2 \fir{1}_{\pi_1}(2) d_{\pi_1} d_1
t^{(1)}_{\pi_1 \pi_2} d_{\pi_1}
\fsai{1}{\pi_1}(2) d_{\pi_2}
\fsai{1}{\pi_2}(1)\nn\\
+ \fir{1}_{\pi_2}(1) d_{\pi_2} d_2 \fir{1}_{\pi_1}(2) d_{\pi_1} d_1
t^{(1)}_{\pi_1 \pi_2} d_{\pi_2}
\fsai{1}{\pi_2}(2) d_{\pi_1}
\fsai{1}{\pi_1}(1)
\label{eq:pipi}
\eea
\end {claim}

For a proof of this result see Appendix~\ref{ap-calcdet1}.

We never intended to include the possibility of $\pi-\pi$
interaction in our equations, since we think it is too small to make a
significant difference to the calculation. This is a reasonable
approximation, since in a model with only pions and nucleons two pions may
only interact via the exchange of
nucleon-anti-nucleon pairs. Consequently the lowest-order diagrams for
$\pi-\pi$ scattering in a model with a $\pi NN$ vertex and a $\pi-N$
contact term  are the two diagrams shown in Figure \ref{fig-box}. When
these diagrams are used in (\ref{eq:pipi}) they produce a $N \bar{N} N N$
intermediate state. Considering that we are already neglecting explicit
$\pi \pi N N$ states in the calculation, and that we are also neglecting
processes involving mesons other than the pion, e.g. the rho meson, the
error made in ignoring diagrams such as those represented by
Eq.~(\ref{eq:pipi})
is insignificant, and consequently from now on we neglect the expression
(\ref{eq:pipi}) as a possible source of double-counting.

Therefore the sum of all diagrams in $c_1 \cap \cfi{N2'}$ is:
\bea
\fir{1}_{\pi'} (1) d_{\pi'} \fir{1}_{\pi} (2) d_1
\fsai{1}{\pi'} (2) d_1 d_\pi \tilde{t}_{\pi N}^{(1)} (2) d_2
d_\pi \fsai{1}{\pi}(1)\nn\\
+ \fir{1}_{\pi_2} (1) d_{\pi_2} \fir{1}_{\pi_1}(2) d_1 t_{\pi_2
N}^{(1)} (2) d_1 d_{\pi_1} \fsai{1}{\pi_1}(2) d_{\pi_2} d_2
\fsai{1}{\pi_2} (1).
\eea

\subsubsection {$c_2$}

The term $c_2$ has one possible cut involving $N2'$ and two
internal lines, as shown in Figure \ref{fig-secondterm}. This diagram
shows that any portion of $\tnni{1}$ which is two-particle reducible in
the $N1' \leftarrow N1 + N2 + N2'$-channel will, when placed in the
expression for $c_2$, lead to diagrams which are in $c_2 \cap \cfi{N2'}$.

Consequently, we need to identify all $N1' \leftarrow N1 + N2 + N2'$-channel
2PR portions of $\tnni{1}$. In order to do this write:
\be
\tnni{1}=\tnni{2} + \tnni{2} d_1 d_2 \tnni{1}.
\label {eq:BSNN}
\ee
$\tnni{2}$ is then split into two pieces: the $t$-channel 1PI part, which
is denoted by $\titnni{2}$, and the $t$-channel one-particle reducible
part, which is clearly just the one-pion exchange potential. Therefore
we obtain:
\be
\tnni{2}=v_{OPE} + \titnni{2},
\ee
(note that this is exactly the alternative decomposition given above in
Eq.~(\ref{eq:altdecomp})) leading to:
\be
\tnni{1}=\fir{1}(2) d_\pi \fai{1}(1) + \titnni{2} + \fir{1}(2)
d_\pi \fai{1}(1) d_1 d_2 \tnni{1} + \titnni{2} d_1 d_2 \tnni{1}.
\label {eq:tnndecomp}
\ee

Examination of the first term on the right-hand side of this
equation shows that if
$\fir{1}(2)$ is two-particle reducible in the $N' \leftarrow N \pi'$
channel then this part of $\tnni{1}$ will be 2PR in the $N1'
\leftarrow N1 + N2 + N2'$-channel. Use of the LICL then shows that:
\be
\fir{1}_{\pi'}(1) d_2 d_{\pi'} \fir{2}_{\pi}(2) d_1 d_\pi t_{\pi
N}^{(1)}(2) d_\pi \fsai{1}{\pi}(1) d_2 \fsai{1}{\pi'}(1),
\label {eq:c21}
\ee
is in $c_2 \cap \cfi{N2'}$ and hence in $D_5$.

Furthermore, the third term on the right-hand side of
(\ref{eq:tnndecomp}) is completely 2PR in the $N1' \leftarrow N1 + N2 +
N2'$-channel and thus the diagram:
\be
\fir{1}_{\pi}(1) d_2 d_\pi \fir{1}_{\pi'}(2) d_{\pi'}
\fsai{1}{\pi'}(1) d_1 d_2 \tnni{1} d_2 \fsai{1}{\pi}(1),
\label{eq:c22}
\ee
is in the set $c_2 \cap \cfi{N2'}$.

Now consider the second and fourth terms of the right-hand side
of Eq.~(\ref{eq:tnndecomp}). The only way in
which these terms lead to portions of $\tnni{1}$ which are 2PR
in the $N1' \leftarrow N1 + N2 + N2'$-channel is if $\titnni{2}$ is
itself 2PR in the same channel. Arguments given in Appendix
\ref{ap-calcdet2} show that the only contributions of such diagrams to
$c_2 \cap C_5^{\{N2'\}}$ are:
\be
\fir{1}_{\pi_1}(2) d_{\pi_1} \fir{1}_{\pi_2}(1) d_2 t_{\pi_1 N}^{(1)}(1)
d_2 d_{\pi_2} \fsai{1}{\pi_2}(1) d_1 d_{\pi_1}
\fsai{1}{\pi_1}(2),
\label{eq:c23}
\ee and:
\be
\fir{1}_{\pi_1}(2) d_1 d_{\pi_1} \fir{1}_{\pi_2}(1) d_{\pi_2} d_2
\titnni{1} d_2 \fsai{1}{\pi_2}(1) d_1 d_{\pi_1}
\fsai{1}{\pi_1}(2),
\label{eq:c24}
\ee where $\titnni{1}$ is not only one-particle irreducible in the
$s$-channel, but also 1PI in the $t$-channel.

Therefore, the parts of $c_2$ which are also in $\cfi{N2'}$ and
so are included in $D_5$ are the diagrams corresponding to
Eqs. (\ref{eq:c21}), (\ref{eq:c22}), (\ref{eq:c23}) and (\ref{eq:c24}).

\subsubsection {$c_3$}

Examination of the diagram representing this term shows that two
three-cuts involving $N2'$ are possible, as shown in Figure
\ref{fig-term3}. The possibility of these three-cuts means that
any diagram contributing to $c_3$ which contains a part of the
$NN$ t-matrix $\tnni{1}$ which is:
\begin{enumerate}
\item 1PR in the $t$-channel;

\item 1PI in the $t$-channel but 2PR in the $N1' \leftarrow N1 + N2 +
N2'$-channel,
\end {enumerate}
is in $c_3 \cap \cfi{N2'}$.

As far as possibility 1 goes, substituting the 1PR piece of $\tnni{1}$,
$v_{OPE}$, into the expression (\ref{eq:c3}) for $c_3$ indicates that
the term: \be
\fir{1}_{\pi_1}(1) d_2 d_{\pi_1} \fir{1}_{\pi_2}(2) d_{\pi_2}
\fsai{1}{\pi_2}(1) d_1 \fsai{1}{\pi_1}(2),
\label {eq:c31}
\ee
is in $c_3 \cap \cfi{N2'} \subset D_5$.

We also showed above and in Appendix \ref{ap-calcdet2}
 that the $N1' \leftarrow N1 + N2 +
N2'$-channel 2PR parts
of the $t$-channel 1PI $NN$ t-matrix are:
\begin{enumerate}
\item $\fir{1}(2) d_\pi \fai{1}(1) d_1 d_2 \tnni{1}$;

\item $\fir{1}(2) d_1 d_\pi \tilde{F}^{(2) \dagger}$;

\item $\fir{1}(2) d_1 d_\pi \tilde{F}^{(2) \dagger} d_1 d_2
\tnni{1}$;
\end{enumerate}
where the amplitude $\tilde{F}^{(2) \dagger}$ is defined in
Appendix~\ref{ap-calcdet2}. However, when these three diagrams are
substituted into the expression for $c_3$ only diagram 1 and the
portion: \be \fir{1}(2) d_1 d_\pi \tnni{1} d_2 \fai{1}(1)
\ee
of diagram 2 produce diagrams which are in $c_3 \cap
\cfi{N2'}$. This is because all the other diagrams produced by
the substitution of $N1' \leftarrow N1 + N2 +
N2'$-channel 2PR parts of
$\titnni{1}$ into $c_3$ will only be contained in $\cfi{N2'}$ if a
three-body force is included in the calculation.

Therefore the contents of $c_3 \cap \cfi{N2'}$ are the diagram
(\ref{eq:c31}), and the diagrams
\be
\fir{1}_{\pi}(1) d_\pi d_2 \fir{1}_{\pi'}(2) d_{\pi'}
\fsai{1}{\pi'}(1) d_1 d_2 \tnni{1} d_1 \fsai{1}{\pi}(2)
\label{eq:c32}
\ee
and:
\be
\fir{1}_{\pi_1}(1) d_2 \fir{1}_{\pi_2}(2) d_1  d_{\pi_1} \tnni{1}
d_{\pi_2} d_2 \fsai{1}{\pi_2}(1) d_1 \fsai{1}{\pi_1}(2).
\label{eq:c33}
\ee

\subsubsection {$c_4$}

If the three-cut
drawn in Figure \ref{fig-Fourthtermcut} may be made
then the diagram representing $c_4$ will also be in
$\cfi{N2'}$.

Therefore diagrams which sum to:
\be
\fir{2}(1) d_2 d_\pi \ttpni{1}(1) d_\pi \fsai{1}{2PR}(2),
\label{eq:c4}
\ee
where $\fsai{1}{2PR}(2)$ is the $N' \leftarrow N + \pi$-channel 2PR part
of the 1PI $\pi NN$ vertex, are in $c_4 \cap \cfi{N2'}$ and so are
in $D_5$.

\subsubsection {$c_5$}

Examination of the diagram representing the sum of $c_5$ shows
that an $s$-channel three-cut involving $N2'$ cannot be produced
by cutting lines beyond the first interaction (see
Fig.~\ref{fig-fifthterm}). Therefore, the
structure of $U^{(2)}_{2 \beta}$ and the amplitudes appearing
after $U_{2 \beta}^{(2)}$ are irrelevant to the calculation of
$c_5 \cap \cfi{N2'}$.

Now an $s$-channel three-cut involving $N2'$ will be possible on
this diagram if:
\begin{enumerate}
\item The $\pi-N$ t-matrix $\tpni{1}$ used immediately before
the pion absorption is $u$-channel 1PR;

\item The $u$-channel 1PI part of that $\tpni{1}$ is
$N' \leftarrow N + \pi + \pi'$-channel 2PR.
\end{enumerate}

Using the above results for the $u$-channel 1PR part of $\tpni{1}$ and
the $N' \leftarrow N + \pi +
\pi'$-channel 2PR part of the $u$-channel 1PI
$\tpni{1}$, but ignoring those diagrams which are only in $\cfi{N2'}$
if a three-body force is included in the calculation, shows that:
\bea
c_5 \cap \cfi{N2'}&=&\sum_{\beta i} \fir{1}_{\pi'}(1) d_2 d_{\pi'}
\fir{1}_{\pi}(2)  d_1 \fsai{1}{\pi'}(2) (1 + d_1 d_\pi
\tpni{1}(2)) \nn\\
&\times& d_1 d_\pi U_{2 \beta}^{(2)} \gtb \tit{1}(\beta)
d_\beta^{-1} \bd{\beta}{i} d_{\bar{i}} d_\pi \fai{1}(i).
\label{eq:c51}
\eea

\subsubsection {$c_6$}

As was the case for $c_5$, we observe  (see Fig.~\ref{fig-sixthterm})
that an $s$-channel three-cut involving $N2'$ cannot be
produced by cutting lines beyond the first interaction.
Therefore, the structure of $U^{(2)}_{3 \beta}$ and the
amplitudes appearing after $U_{3 \beta}^{(2)}$ are irrelevant to
the calculation of $c_6 \cap \cfi{N2'}$.

Then an $s$-channel three-cut involving $N2'$ is possible in
Fig.~\ref{fig-sixthterm} if:

\begin{enumerate}
\item The 1PI $N-N$ t-matrix $\tnni{1}$ used immediately before
the pion absorption is $t$-channel 1PR.

\item The $t$-channel 1PI $NN$ t-matrix, $\titnni{1}$ is  $N1'
\leftarrow N1 + N2 + N2'$-channel 2PR.
\end{enumerate}

Using the previous results for these portions of the $NN$ t-matrix and
again ignoring those parts of $c_6$ which are only also in $\cfi{N2'}$
if a three-body force is included in the calculation we find:
\bea
c_6 \cap \cfi{N2'}&=&\sum_{\beta i} \fir{1}_{\pi}(1) d_2 d_{\pi}
\fir{1}_{\pi'}(2) d_{\pi'} \fsai{1}{\pi'}(1) (1 + d_1 d_2
\tnni{1})\nn\\
&\times&  d_1 d_2 U_{3 \beta}^{(2)} \gtb \tit{1}(\beta)
d_\beta^{-1} \bd{\beta}{i} d_{\bar{i}} d_\pi \fai{1}(i).
\eea

\subsubsection {Overall result for $D_5$}

Combining the results for the intersections $c_a \cap
\cfi{N2'}$, $a=1,\ldots,6$, as per Eq.~(\ref{eq:sumint}) reveals
that:
\bea
D_5&=&\sum_{\beta i} \fir{1}_{\pi'}(1)
d_2 d_{\pi'} \fir{1}_{\pi}(2)  d_1 \fsai{1}{\pi'}(2) (1 + d_1
d_\pi \tpni{1}(2)) d_1 d_\pi U_{2 \beta}^{(2)} \gtb \tit{1}(\beta)
d_\beta^{-1} \bd{\beta}{i} d_{\bar{i}} d_\pi \fai{1}(i)\nn\\
&+& \sum_{\beta i} \fir{1}_{\pi}(1) d_2 d_{\pi}
\fir{1}_{\pi'}(2) d_{\pi'} \fsai{1}{\pi'}(1) (1 + d_1 d_2
\tnni{1})  d_1 d_2 U_{3 \beta}^{(2)} \gtb \tit{1}(\beta)
d_\beta^{-1} \bd{\beta}{i} d_{\bar{i}} d_\pi \fai{1}(i)\nn\\
&+& \fir{2}(1) d_2 d_\pi \ttpni{1}(1) d_\pi \fsai{1}{2PR}(2)\nn\\
&+& \fir{1}_{\pi'}(1) d_2 d_{\pi'} \fir{2}_{\pi}(2) d_1 d_\pi
t_{\pi N}^{(1)}(2) d_\pi \fsai{1}{\pi}(1) d_2
\fsai{1}{\pi'}(1)\nn\\
&+& B + X + D_1 + D_2 + \tilde{D}_{\pi}\nn\\
&+& \fir{1}_{\pi}(1) d_\pi d_2 \fir{1}_{\pi'}(2) d_{\pi'}
\fsai{1}{\pi'}(1) d_1 d_2 \tnni{1} d_1 \fsai{1}{\pi}(2)\nn\\
&+& \fir{1}_{\pi}(1) d_\pi d_2 \fir{1}_{\pi'}(2) d_{\pi'}
\fsai{1}{\pi'}(1) d_1 d_2 \tnni{1} d_1 \fsai{1}{\pi}(1)\nn\\
&+& \fir{1}_{\pi'}(1) d_{\pi'} \fir{1}_{\pi}(2) d_1
\fsai{1}{\pi'}(2) d_1 d_\pi
\ttpni{1}(2) d_2 d_\pi \fai{1}(1),
\label{eq:D5res}
\eea
where:
\be
B=\fir{1}_{\pi_1}(1) d_2 \fir{1}_{\pi_2}(2) d_{\pi_1} d_{\pi_2}
\fsai{1}{\pi_2}(1) d_1 \fsai{1}{\pi_1}(2),
\ee
represents crossed two-pion exchange; $X$ is given by:
\be
X=\fir{1}_{\pi_1}(1) d_2 \fir{1}_{\pi_2}(2) d_1  d_{\pi_1}
\tnni{1} d_{\pi_2} d_2 \fsai{1}{\pi_2}(1) d_1 \fsai{1}{\pi_1}(2),
\ee
the terms $D_i$ are:
\be
D_i=\fir{1}_{\pi_2}(1) d_{\pi_2} d_2 \fir{1}_{\pi_1}(2) d_1
d_{\pi_1} t_{\pi_i N}^{(1)} (i) d_i^{-1} d_1 d_{\pi_i}
\fsai{1}{\pi_1}(2) d_2
\fsai{1}{\pi_2}(1),
\ee
and $\tilde{D}_{\pi}$ is given by:
\be
\tilde{D}_{\pi}=\fir{1}_{\pi_2}(1) d_{\pi_2} d_2
\fir{1}_{\pi_1}(2) d_1 d_{\pi_1} \tilde{T}_{NN}^{(1)} d_1
\fsai{1}{\pi_1}(2) d_2 \fsai{1}{\pi_2}(1).
\ee

At this stage we must make a decision. Upon substitution of
the expression for $D_5$ into (\ref{eq:T22}) many of the terms in
$D_5$ can be subtracted from either the $N-N$ t-matrix or the $\pi-N$
t-matrix. In what follows we choose to subtract them from the $\pi-N$
t-matrix, as we did above for the t-matrices in $\bfai{2}$. This has the
effect that many of the ``potentials" generating the $\pi-N$ t-matrices,
which
are, by definition, 2PI in the $s$-channel, also become 1PI in the $u$-channel.
In other words, the resultant input $\pi N$ t-matrix appearing at some
points in
the equations has the crossed diagram removed from its ``potential".
However, we
stress that it would be equally valid to modify the $NN$ t-matrix. Were this to
be done the modified $NN$ ``potential" would be 2PI in the $s$-channel
{\em and}
1PI in the $t$-channel. That is, if we chose to proceed in this way the input
$NN$ t-matrix which appears at some points in the equations will not contain
the one-pion
exchange contribution in the driving term of its Bethe-Salpeter
equation.

As we shall see further below, the fact that we choose to modify the $\pi-N$
t-matrix means that the $\pi NN$ form factor in the one-pion exchange
part of $\tnni{2}$ needs to be adjusted. By contrast, if we chose to
modify the $NN$ t-matrix it may be shown that this change to the $\pi
NN$ form factor in one-pion exchange is unnecessary. This result is
merely a specific example of the wider truth that any adjustments which
are made in  the pion production/absorption sector have significant
implications for the $NN \rightarrow NN$ sector of the theory.

\subsection {Equation for $\tnni{2}$}

Having made this decision we may substitute
eqs.(\ref{eq:bfa2}), (\ref{eq:D4res}) and (\ref{eq:D5res}) into
Eq.~(\ref{eq:T22}), and simplify using the AGS equations in order
to obtain:
\bea
\tnni{2}&=& \fir{2}(1) d_\pi \tfai{1}(2) + \tilde{C}_5^d \nn\\
&+&\sum_{j \alpha i} \fir{1}(j) \bd{j}{\alpha}
d_{\bar{j}} d_\pi \titit{1}(\alpha) d_\alpha^{-1} d_{\bar{i}}
d_\pi \bd{\alpha}{i} \fai{1}(i) \nn\\
&+&\sum_{j \alpha \beta i}
\fir{1}(j) \bd{j}{\alpha} d_{\bar{j}} d_\pi
\tilde{t}^{(1) \dagger}(\alpha) d_\alpha^{-1} \gtb U_{\alpha
\beta}^{(2)}
\gtb \tit{1}(\beta) d_\beta^{-1} d_{\bar{i}} d_\pi
\bd{\beta}{i} \fai{1}(i)\nn\\
&-&D_1 -D_2 - \tilde{D}_\pi-X-B,
\label {eq:C5}
\eea
where:
\be
\ttpna{1}=v^R + v^R d_N d_\pi t^{(1)},
\ee
and:
\be
\ttpni{1}=v^R + t^{(1)} d_N d_\pi v^R,
\ee
are both one-particle irreducible in the $s$ and $u$-channels, but
differ in that $\ttpni{1}$ has the crossed term subtracted from
the first interaction, while $\ttpna{1}$ has it
subtracted from the last interaction. By
contrast: \be
\tilde{\tilde{t}}^{(1)} (i)=\tilde{\tilde{t}}_{\pi N}^{(1)}=v^R + v^R d_N
d_\pi v^R + v^R d_N d_\pi t_{\pi N}^{(1)} d_N d_\pi v^R,
\ee
for $i=1,2$, has the crossed term subtracted from both the first
and the last term. Meanwhile:
\be
\tilde{\tilde{t}}^{(1)} (3)=\tnni{1}.
\ee
and $\tilde{C}_5^d$ is that portion of the sum of class
$C_5$ which contributes to the dressing of one-pion exchange and
is given by:
\bea
\tilde{C}_5^d&=&\fir{2}(2) d_1 d_\pi \ttpni{1}(2) d_\pi
\fai{1}(1)\nn\\
&&+\fir{2}(1) d_1 d_\pi \ttpni{1}(1) d_\pi
\tfai{1}(2)\nn\\
&&-\fir{2}_{\pi}(2) d_1 d_\pi t_{\pi N}^{(1)}(2) d_\pi
\fir{1}_{\pi'}(1) d_2 \fsai{1}{\pi}(1) d_2 d_{\pi'}
\fsai{1}{\pi'}(1).
\eea

Formal manipulations now show that:
\be
\tilde{C}_5^d + f^{(2)}(1) d_\pi \tfai{1}(2)= V^*_{OPE} + D_{OPE},
\label{eq:C5dt}
\ee
where:
\bea  D_{OPE} &=& \fir{1}_{\pi_2}(1) d_{\pi_2} d_2 \fir{1}_{\pi_1}(2) d_1
d_{\pi_1} \fir{1}_{\pi_3}(2) d_{\pi_3} \fsai{1}{\pi_3}(1) d_1
\fsai{1}{\pi_1}(2) d_2
\fsai{1}{\pi_2}(1);\\
V^*_{OPE} &=& f^{(1) *} (1) d_\pi {f^{(1) *}}^{\dagger}(2);
\eea
with:
\be
f^{(1) *}=f^{(2)} + f^{(2)} d_N d_\pi \ttpni{1},
\ee
the vertex which is obtained if the crossed diagram is eliminated
(in the manner prescribed above) from the $\pi-N$ t-matrix used
to generate the dressed $\pi NN$ vertex. Note that, using the
definition of $\tilde{t}_{\pi N}$, Eq.~(\ref{eq:tildet2}), we
find:
\be
f^{(1) *}=f^{(1)}-\fir{1}_{\pi'} d_N \fir{1}_{\pi} d_N d_{\pi'}
\fsai{1}{\pi'}
\label{eq:f*}
\ee
as depicted in Figure \ref{fig-Newvertex}. Note that in order to obtain
the result (\ref{eq:C5dt}) we have had to use the
time-dependent perturbation theory identities (\ref{eq:verteq}) and
(\ref{eq:tdpt1}).

Resubstituting the result (\ref{eq:C5dt}) into Eq.~(\ref{eq:C5})
yields the final result for $\tnni{2}$:
\bea
\tnni{2}&=&V_{OPE}^* - D_1 - D_2 - D_\pi - X - B \nn\\
&+& \sum_{j \alpha i} \fir{1}(j) \bd{j}{\alpha} d_{\bar{j}} d_\pi
\tilde{\tilde{t}}^{(1)}(\alpha) d_\alpha^{-1} d_{\bar{i}} d_\pi \bd{\alpha}{i}
\fai{1}(i)\nn\\
&+& \sum_{j \alpha \beta i} \fir{1}(j)
\bd{j}{\alpha} d_{\bar{j}} d_\pi
\tilde{t}^{(1) \dagger}(\alpha) d_\alpha^{-1} \gtb U_{\alpha
\beta}^{(2)} \gtb \tilde{t}^{(1)}(\beta) d_\beta^{-1} d_{\bar{i}}
d_\pi \bd{\beta }{i} \fai{1}(i),
\label {eq:finalT2}
\eea
where:
\be
D_\pi=\tilde{D}_\pi+D_{OPE}=\fir{1}_{\pi_2}(1) d_{\pi_2} d_2
\fir{1}_{\pi_1}(2) d_{\pi_1} d_1 T_{NN}^{(1)} d_1
\fsai{1}{\pi_1}(2) d_2 \fsai{1}{\pi_2}(1).
\ee

The amplitude $\tnni{2}$ is the nucleon-nucleon $s$-channel two-particle
irreducible interaction which is to be used in the Bethe-Salpeter equation for
$NN$ scattering in this theory, Eq.~(\ref{eq:BSNN1}). Hence it plays the role
of
the nucleon-nucleon ``potential" in this work. However, rather than merely
being
the sum of a few Feynman diagrams $\tnni{2}$ is the sum of {\em all} 2PI
Feynman
diagrams with one explicit pion. Consequently it includes the full explicit
coupling to the
$\pi NN$ channel.

If this result is compared to the equation for $\tnni{2}$ obtained in the
derivation of the standard $NN-\pi NN$
equations~\cite{Th73,Mi76,MK77,Ri77,TR79,AM79,AM80,AM81,AM83,AB80,AB81}
a number
of differences are seen:
\begin{enumerate}
\item The one-pion exchange part of the ``potential" has been replaced by the
modified ``potential":
\be
\bar{V}=V^*_{OPE} - D_1 - D_2 - D_\pi - X - B,
\label{eq:Vbar}
\ee
that actually contains (for reasons to be discussed below) the diagrams $D_1$,
$D_2$, $D_\pi$, $X$ and $B$ which involve multiple pion exchanges.

\item In the two terms describing pion rescattering (i.e. the last two terms
on the right-hand side of Eq.~(\ref{eq:finalT2})) the $\pi-N$
t-matrix has been modified. In the first term the usual 1PI $\pi-N$ t-matrix
has
been supplanted by the t-matrix $\titit{1}$, in which the crossed diagram has
been removed from both the first and the last pion-nucleon ``potential".
Likewise, in the second term the operators $\tit{1}$ and
$\tait{1}$ govern, respectively, the first/last pion rescattering after/before
pion emission/absorption.
\end{enumerate}

All of these changes are the result of our desire to eliminate double-counting
from the equations. Taking the second set of changes first, in Section
\ref{sec-F2adj} it was shown that either the $\pi N$ or $NN$ amplitude
immediately following a pion emission needed to be modified in order to avoid
the over-counting of diagrams involving the process (\ref{eq:D4res1}). The
changes to these $\pi-N$ t-matrices in Eq.~(\ref{eq:finalT2}) are made for
exactly the same reason. The effect is to place constraints on these
sub-amplitudes in channels other than the $s$-channel. As discussed in
Section~\ref{sec-Taylorrev}, such additional constraints are necessary because
in a time-dependent perturbation theory diagram constraints on the $s$-channel
cut-structure of sub-amplitudes are not enough to constrain the overall
$s$-channel cut-structure of the amplitude.

Turning now to the first set of changes, we divide our discussion into two
halves. Firstly, let us consider the one-pion exchange piece of $\bar{V}$,
$V^*_{OPE}$. A very interesting consequence of our work here is that the
one-pion exchange ``potential" $V^*_{OPE}$ contains a different vertex than
that
used in the pole part of the $\pi-N$ t-matrix $\tpni{0}$. This is necessary
because the inclusion of the full $\pi NN$ vertex in the one-pion exchange
potential leads to double-counting with respect to other terms in $\tnni{2}$.
(See Fig.~\ref{fig-Vertexovercount} for examples of such diagrams.) These other
terms are included in the final two terms of Eq.~(\ref{eq:finalT2}). Therefore,
they only arise when the $NN \pi$ cut structure of the amplitude $\tnni{2}$ is
considered.  This is a manifestation of the fact that the amount of
double-counting to be eliminated is entirely dependent upon which
unitarity cuts are opened in the analysis. Our above discovery that certain
types of double-counting do not arise when a three-body force is excluded from
the calculation is another manifestation of the same effect.
Therefore we find that whether or not the vertex used for the one-pion exchange
piece of $\tnni{2}$ needs to be adjusted depends on whether:
\begin{enumerate}
\item the $\pi NN$ unitarity cut is opened or not;

\item the subtraction terms are included in $t_{\pi N}$ or $T_{NN}$ (see
discussion in Section~\ref{sec-calcD5}).
\end{enumerate}

Secondly, the other subtractions in $\bar{V}$ (i.e. $D_1$, $D_2$, $D_\pi$, $B$
and $X$) are also required in order to correct for  the overcounting of
diagrams
in the last two terms of Eq.~(\ref{eq:finalT2}). (Note that once again it is
the
opening of the $\pi NN$ cut which has led to double-counting.)  As is shown in
Figs.~\ref{fig-Vbar2} and \ref{fig-Vbar1} $D_1$, $D_2$, $D_\pi$, $B$ and $X$
are
all included in this term in more than one place. Consequently these five
diagrams will be double-counted unless some adjustment to the equation for
$T_{NN}^{(2)}$ is made. Because we wish to obtain a set of coupled equations we
choose to make this adjustment in the $NN$ ``potential", rather than in the
pion absorption/production channel.

The necessity of eliminating double-counting from the theory is therefore
forcing either:
\begin{enumerate}
\item The modification of the amplitudes appearing within the
equations, or

\item The inclusion of explicit subtraction terms to compensate for the
over-counting.
\end{enumerate}

If option 1 is pursued the result is that different amplitudes appear at
different places within the theory: for instance, we have four $\pi N$
amplitudes in the equation~(\ref{eq:finalT2}). If option 2 is taken then the
presence of extra subtraction terms means that the driving term of the equation
takes on a far more complicated form, since it now involves diagrams with more
than one explicit pion. In either case the original goal of deriving
coupled equations for the $NN-\pi NN$ system which contain only one explicit
pion
and standard 1PI $NN$  and $\pi N$ amplitudes has been defeated. If such
equations were to be  derived, as was done by AM and AB, certain diagrams would
need to be (incorrectly) included in {\em two} (or more) places in the theory.
That these diagrams appear {\em once and only once} in the diagrammatic
expansion is a direct consequence of the indeterminacy of the  time-ordering of
interaction vertices on different particles in a time-dependent field theory.

\section {Derivation of coupled equations for the $NN-\pi NN$ system}

\label {sec-coupled}

Our next task is to obtain double-counting-free coupled integral equations for
the amplitudes which describe the processes:
\begin{equation}
\left. \begin{array}{rr} N_1 + N_2\\ (N_2 \pi) + N_1\\ (N_1 \pi) + N_2\\ (N_1
N_2) + \pi
\end{array} \right \}
 \longrightarrow
\left \{
\begin{array}{ll}  N_1 + N_2\\  (N_2 \pi) + N_1\\  (N_1 \pi) + N_2\\
(N_1 N_2) + \pi
\end{array} \right.
\end{equation}
This will be achieved in two stages: firstly, integral
equations governing the amplitudes which describe the reactions
$NN \rightarrow NN$, $NN \rightarrow NN \pi$, $NN \pi \rightarrow
NN$, $NN \pi \rightarrow NN \pi$ will be found. Once this is
accomplished the residues of these amplitudes at the appropriate
poles will be taken, in order to derive equations connecting the
two-fragment amplitudes for the above processes.

\subsection {Deriving integral equations for the
two and three-body amplitudes}

In this section the first of these two steps is performed: the
derivation of integral equations for the two and three-body
amplitudes. We have, of course, already begun this task by
deriving equations for the amplitudes $\tnni{1}$, $\tnni{2}$ and
$\bfai{2}$. But, equations are now needed, not only for these
amplitudes, but also for the connected $NN \pi \rightarrow NN \pi$
amplitude $M^{(1)}$, the connected $NN \rightarrow NN \pi$ amplitude
$\bfai{1}$, and the connected $NN \pi \rightarrow NN$ amplitude,
$F^{(1)}$. (Note that since all the amplitudes under
consideration involve a two-nucleon initial state, by conservation
of nucleon number they are automatically one-particle irreducible.)

\subsubsection   {$\bfai{1}$ and $F^{(1)}$}

Firstly, consider the connected 1PI $NN \rightarrow NN \pi$
amplitude $\bfai{1}$. Since the parameters of the Taylor method
for this case are $n=3$, $m=2$ and $r=2$ no double-counting can
arise and so the classification-of-diagrams technique may be
applied to this amplitude without adaptation. This process produces
the equation:
\be
\bfai{1}=\bfai{2} (1 + d_1 d_2 \tnni{1}) + \sum_j \fai{1}(j) d_1
d_2 \tnni{1},
\label {eq:tbfa1}
\ee
where the first term is generated by $C_1$, the second by
$C_2$, and the third by $C_5$. Classes $C_3$ and $C_4$ are
empty, since, as has already been observed above, all one-to-one
amplitudes are zero.

Substituting the expression obtained for $\bfai{2}$ given in
Eq.~(\ref{eq:bfa2}) then produces:
\bea
\bfai{1}=\sum_j \fai{1}(j) d_1 d_2 \tnni{1} + \left[ \sum_{\alpha
i} \tit{1}(\alpha) d_\alpha^{-1} d_{\bar{i}} d_\pi
\bd{\alpha}{i} \fai{1}(i) \right.\nn\\ + \left.
\sum_{\alpha \beta i} t^{(1)} (\alpha) d_{\alpha}^{-1}
\gtb U_{\alpha \beta}^{(2)} \gtb \tit{1}(\beta) d_{\beta}^{-1}
\bd{\beta}{i} d_{\bar{i}} d_\pi \fai{1}(i) \right] (1+d_1 d_2
\tnni{1}).
\label {eq:bfa1}
\eea
Taking the adjoint of Eq.~(\ref{eq:bfa1}) then gives an
expression for the 1PI $NN \pi \rightarrow NN$ amplitude,
$F^{(1)}$, in terms of two-body t-matrices and dressed $\pi NN$
vertices.

\subsubsection {$M^{(1)}$}

In this case $m=n=3$ and $r=2$, and therefore Taylor's original
method may again be applied, giving:
\be M^{(1)}=M^{(2)} + \bfai{2} d_1 d_2 F^{(1)} + \bfai{2} d_1 d_2
\left[ \sum_{i=1}^2 \fir{1}(i) d_i^{-1} \right] +
\left[ \sum_{j=1}^2 \fai{1}(j) d_j^{-1} \right] d_1 d_2 F^{(1)},
\label{eq:M1}
\ee where the four terms come from, respectively, $C_1$, $C_2$,
$C_3$ and $C_5$. Note that $C_4$ is empty. Note also that we have
made use of Eq.~(\ref{eq:f0}) for $f^{(0)}$ here.

In order to derive an expression for $M^{(1)}$ in
terms of two-body t-matrices and dressed $\pi NN$ vertices an
equation for $M^{(2)}$ must be derived. In the case of $M^{(2)}$
the parameters of the Taylor method are $n=m=r=3$ and so one might
expect that, since $n=r$, double-counting will occur in $C_4$ and
$C_5$. However, since all of the particles are fully dressed no
three-cut may intersect more than one line from the final state and
so in every diagram in $C_4$ and $C_5$ we have $s_f=1$.
Consequently, the condition, $n > 2s_f$ is always satisfied and so
it is guaranteed that no double-counting occurs even if Taylor's
original method is used. When the original Taylor method is applied
to the amplitude $M^{(2)}$ it yields: \be
M^{(2)}=M^{(3)}+\left\{ \left[ M^{(3)} + \sum_\alpha t^{(2)}
(\alpha) d_\alpha^{-1} \right] d_1 d_2 d_\pi \left[M^{(2)} +
\sum_{\beta} t^{(1)}(\beta) d_\beta^{-1}\right]\right\}^{(c)},
\ee
where the superscript ${}^{(c)}$ indicates that the amplitude
resulting from the multiplication must be connected.

Since we have consistently ignored three-body $NN \pi$ forces
throughout this calculation, we set $M^{(3)}$ to zero, thus
producing the equation:
\be
M^{(2)}=\sum_{\alpha \beta} t^{(2)}(\alpha)
d_\alpha^{-1} d_1 d_2 d_\pi \bd{\alpha}{\beta} t^{(1)} (\beta)
d_\beta^{-1} + \sum_{\alpha} t^{(2)}(\alpha) d_\alpha^{-1} \gtb
M^{(2)}.
\ee
This equation may then be formally solved, in order to yield:
\be
M^{(2)}=\sum_{\alpha \beta} t^{(1)}(\alpha)
d_\alpha^{-1} d_1 d_2 d_\pi U^{(2)}_{\alpha \beta} d_1 d_2
d_\pi t^{(1)} (\beta) d_\beta^{-1},
\label{eq:M2}
\ee
where $U^{(2)}_{\alpha \beta}$ are the covariant AGS amplitudes,
which obey Eqs.~(\ref{eq:AGS1})
and (\ref{eq:AGS2}).

Using Eqs.~(\ref{eq:bfa2}), (\ref{eq:bfa1}) and (\ref{eq:M2}) in
Eq.(\ref{eq:M1}) an expression for $M^{(1)}$ in terms of the $\pi N$
and $NN$ amplitudes and the dressed $\pi NN$ vertex may be derived.

\subsection {Coupled equations for the $NN \rightarrow NN$
and $NN \pi \rightarrow NN$ reactions}

Eqs.~(\ref{eq:BSNN1}), (\ref{eq:finalT2}), (\ref{eq:bfa1}) and
(\ref{eq:M1}) are equations for the two and three-body
amplitudes of interest. However, our goal is to derive equations
connecting the amplitudes for the collision of two fragments,
rather than equations containing the three-body amplitudes. So
consider firstly the reactions in which a two-fragment final
state that includes a pion is produced from an initial $NN$
state:
\be
N_1 + N_2
\longrightarrow \left \{ \begin{array}{ll}
(N_2 \pi) + N_1\\
(N_1 \pi) + N_2\\
(N_1 N_2) + \pi
\end{array} \right.
\ee
Suppose that the final state of interest is one in which the
particle $\lambda$ is a spectator while the other two
particles are in a bound or resonance state. If the energy of
this final state is $E_\lambda$ then the matrix element
for such a process, $X_{\lambda N}^{D}$, is given
by \cite{Bl59}:
\be
X_{\lambda N}^{D}=(-i) \langle \phi(\lambda)|
\langle \chi_\lambda| d_\lambda^{-1} \gtb \mbox{Res}_{\lambda
\mbox{ pole}}
\gtb d_\lambda^{-1}  {F^{(1)}_D}^\dagger |\psi_D \rangle
\label{eq:Xlndef}
\ee
where $|\phi (\lambda) \rangle$ is the form factor for the
formation of the bound or resonance state; $|\chi_\lambda
\rangle$ and $d_\lambda$ are, respectively, the wave
function and  dressed propagator for a free $\lambda$ particle
and $|\psi_D \rangle$ is a wave function for the initial pair $N_1$
and $N_2$. Note that so far we have treated all particles as
distinguishable, hence the initial state $NN$ wave function must be
the one for distinguishable particles, and the form-factor
$|\phi(\lambda) \rangle$ should not be anti-symmetrized, even if
$\lambda$ represents the channel $(N_1 N_2) + \pi$. Furthermore, we
indicate the fact that the equations for $\bfai{1}$ assume
distinguishable particles by placing the subscript ${}_D$ on
$\bfai{1}$.

In order to derive an equation for $X_{\lambda N}^D$ the
residue of Eq.~(\ref{eq:bfa1}) must be taken at the final state
$\lambda$ pole. It is therefore necessary to make explicit
the analytic structure ``near" this pole of each of the
amplitudes which appear last in the various terms on
the right-hand side of Eq.~(\ref{eq:bfa1}).
To facilitate this we first write:
\be
\tit{1}(\alpha)=\left\{ \begin{array}{ll}
t^{(1)}(\alpha)-t^{(1)}(\alpha) d_N d_\pi
v^X(\alpha) - v^X(\alpha) &\mbox { if $\alpha=1,2$}\\
t^{(1)}(\alpha) &\mbox { if $\alpha=3$}
\end{array}
\right.
\ee
in Eq.~(\ref{eq:bfa1}). Now, if $p_\alpha$ is the
total four-momentum of the two-body sub-system labeled by $\alpha$
then it may be shown that the analytic structure of
$t^{(1)}(\alpha)$ ``near" the
$\alpha$ pole, $p_\alpha^2=M_\alpha^2$, is given by:
\be
t^{(1)}(\alpha) \stackrel{p_\alpha^2 \rightarrow
M_\alpha^2}{\sim}  |\phi(\alpha) \rangle
\frac{i}{p_\alpha^2-M_\alpha^2} \langle \phi(\alpha)|  +
\mbox{(regular terms)}.
\label{eq:tstr}
\ee
When the three-body operator $t^{(1)}(\alpha) d_\alpha^{-1}$ is
considered it is found that if the four-momentum of the spectator
$\alpha$ is $k_\alpha$, the $\alpha$ pole is at the point
$(P-k_\alpha)^2=M_\alpha^2$ where $P$ is the total four-momentum
of the system. Near that point:
\bea
t^{(1)}(\alpha) d^{-1}_\alpha (k_\alpha) &\stackrel{(P-k_\alpha)^2
\rightarrow M_\alpha^2}{\sim}&  d^{-1}_\alpha (k_\alpha)
|\phi(\alpha) \rangle |\chi_\alpha (k_\alpha) \rangle
\frac{i}{(P-k_\alpha)^2-M_\alpha^2} d_\alpha (k_\alpha) \langle
\phi(\alpha)| \langle \chi_\alpha (k_\alpha)| d^{-1}_\alpha
(k_\alpha) \nn\\ &+& \mbox{(regular terms)}.
\eea

The amplitudes other than $t^{(1)}$ which appear last in some term
in Eq.~(\ref{eq:bfa1}) are $v^X$ and $\fai{1}$. Since the former is
the crossed term of the $\pi-N$ potential and the latter is the
dressed $N \rightarrow \pi N$ form factor, neither $v^X$ or
$\fai{1}$ have a pole at any of the points
$(P-k_\lambda)^2=E_\lambda^2$. (The exception to this statement
occurs if $t_{\pi N}^{(1)}$ includes a resonance with the same
quantum numbers as the nucleon. In this case care must be
exercised as the residue may receive a contribution from
$\fai{1}=(1 + t_{\pi N}^{(1)} d_N d_\pi) \fai{2}.$)
Consequently, any term in which a
$v^X$ or $\fai{1}$ is the last amplitude disappears upon the taking
of the residue. It follows that Eqs.~(\ref{eq:Xlndef}) and
(\ref{eq:bfa1}) lead to the following form for $X_{\lambda N}^{D}$:
\be
X_{\lambda N}^{D}=\langle \phi(\lambda)| \langle
\chi_\lambda| d_\lambda^{-1} d_1 d_2 d_\pi T_{\lambda
N}^D|\psi_D
\rangle,
\ee
where the amplitude $T_{\lambda N}^D$ is given
by:
\bea
T_{\lambda N}^D&=&\sum_{j} (1 -
v^X(\lambda) d_\lambda^{-1} \gtb) \bd{\lambda}{j}
\fai{1}(j) d_j^{-1} (1 + d_1 d_2 T_{NN}^D)\nn\\
&+& \sum_{\beta i}
U^{(2)}_{\lambda \beta} \gtb \tit{1}(\beta) d_\beta^{-1}
\bd{\beta}{i} d_{\bar{i}} d_\pi \fai{1}(i) (1 + d_1 d_2 T_{NN}^D).
\label{eq:TlambdaN1}
\eea
Here we have replaced the $NN$ scattering amplitude
$T_{NN}^{(1)}$ by the operator $T_{NN}^D$, since, as discussed
above, the 1PI $NN$ scattering amplitude is the full $NN$
scattering amplitude. The AGS equations may now be used in order
to obtain:
\bea
T_{\lambda N}^D &=& \left[ \sum_j (1 -
 v^X(\lambda) d_\lambda^{-1} \gtb) \bd{\lambda}{j} \fai{1}(j)
d_j^{-1} - \sum_{\alpha j} \bd{\lambda}{\alpha} v^X(\alpha)
d_\alpha^{-1} d_{\bar{j}} d_\pi
\bd{\alpha}{j} \fai{1}(j) \right]\nn\\
& & \times (1 +  d_1 d_2 T_{NN}^D)
+ \sum_{\alpha} \bd{\lambda}{\alpha} t^{(1)} (\alpha)
d_\alpha^{-1} \gtb T_{\alpha N}^D.
\label{eq:TlN}
\eea
Note that this equation is equivalent to that derived by
Avishai and Mizutani (AM) \cite{AM83} and Afnan and
Blankleider (AB) \cite{AB80}, except for the presence of the
term:
\be
-\left[\sum_{\alpha j} v^X(\alpha)
d_\alpha^{-1} d_{\bar{j}} d_\pi \bd{\alpha}{j} \fai{1}(j) \right]
(1+d_1 d_2 T_{NN}^D),
\ee
which has been included in order to remove the double-counting
present when the equations derived by AM and AB are used in a
time-dependent perturbation theory.

The equation derived for $T_{\lambda N}^D$ contains the
amplitude $T_{NN}^D$. Therefore it is now necessary to
derive an equation for $T_{NN}^D$. Recall that
Eqs.~(\ref{eq:BSNN1}) and (\ref{eq:finalT2}) are the results given
by the modified Taylor method for $T_{NN}$. Substituting
(\ref{eq:finalT2}) into (\ref{eq:BSNN1}) yields:
\bea
T_{NN}^D&=&(V^*_{OPE} - D_1 - D_2 - D_\pi - X - B) (1 + d_1 d_2
T_{NN}^D) + \left( \sum_{j \alpha i} \fir{1}(j) \bd{j}{\alpha}
d_{\bar{j}} d_\pi \tait{1} (\alpha) d_\alpha^{-1}
\right) \nn\\  & & \gtb \left[(1 - v^X(\alpha) d_\alpha^{-1})
\gtb \bd{\alpha}{i} \fai{1}(i) d_i^{-1}
+ \sum_{\beta} U^{(2)}_{\alpha \beta}
\gtb
\tit{1}(\beta) d_\beta^{-1} d_{\bar{i}} d_\pi \bd{\beta}{i}
\fai{1}(i) \right] \nn\\
& & (1 + d_1 d_2 T_{NN}^D).
\eea

Using Eq.~(\ref{eq:TlambdaN1}) for $T_{\alpha N}$ then reveals
that: \be
T_{NN}^D=\bar{V} (1 + d_1 d_2 T_{NN}^D) + \sum_{j \alpha}
 \fir{1}(j) \bd{j}{\alpha} d_{\bar{j}} d_\pi
\tait{1} (\alpha) d_\alpha^{-1} \gtb T_{\alpha N}^D,
\label{eq:TNN}
\ee
where $\bar{V}$ is given by Eq.~(\ref{eq:Vbar}).

Recall that in the discussion after Eq.~(\ref{eq:finalT2}) it was
emphasized that the final result for $\bar{V}$ is dependent upon
the treatment of the $NN \leftrightarrow NN \pi$ sector of the
theory. Similarly, since Eq.~(\ref{eq:TNN}) results from a
consistent treatment of the coupled $NN-\pi NN$ system one cannot
simple-mindedly modify this result by, for instance, merely
dropping the second term on the right-hand side. Any
such change must be developed in a coherent way throughout the
theory, rather than imposed in an {\em ad hoc} way on the final
equations.

Because we have used the modified Taylor method and so eliminated
all double-counting the set of four equations given in
Eqs.~(\ref{eq:TlN}) and (\ref{eq:TNN}) are double-counting-free
coupled four-dimensional integral equations for the $NN \rightarrow
\pi NN$ and $NN
\rightarrow NN$ channels.

\subsection {Coupled equations for the $\pi NN \rightarrow
\pi NN$ and $NN \rightarrow NN \pi$ reactions}

We now turn our attention to
the $\pi NN \rightarrow \pi NN$ and $NN \rightarrow \pi NN$
channels, and attempt to derive coupled equations for the
reactions:
\begin{equation}
\left. \begin{array}{rr}
(N_2 \pi) + N_1\\
(N_1 \pi) + N_2\\
(N_1 N_2) + \pi
\end{array} \right \}
 \longrightarrow
\left \{
\begin{array}{ll}
(N_2 \pi) + N_1\\
(N_1 \pi) + N_2\\
(N_1 N_2) + \pi
\end{array} \right.
\label{eq:secreac}
\end{equation}
and:
\begin{equation}
N_1 + N_2 \longrightarrow
\left \{
\begin{array}{ll}
(N_2 \pi) + N_1\\
(N_1 \pi) + N_2\\
(N_1 N_2) + \pi.
\end{array} \right.
\label{eq:secreac2}
\end{equation}

The two-to-two amplitude for any of the reactions
(\ref{eq:secreac}) is found by taking the right and left residue
of $M^{(1)}_D$ at the relevant poles. That is, the matrix element
for a transition from a state in which
particle $\lambda$ is a spectator to a state in
which particle $\mu$ is a spectator is $X_{\lambda \mu}^D$, given
by:
\bea
X_{\lambda \mu}^D &=& (-i)^2\langle \phi(\lambda)| \langle
\chi_\lambda| d_\lambda^{-1} \gtb \mbox{Res}_{\lambda \mbox{ pole}}
\gtb d_\lambda^{-1}  M_D^{(1)}  d_\mu^{-1} \gtb
\nn\\
&&\mbox{Res}_{\mu
\mbox{ pole}} \gtb d_\mu^{-1} |\phi(\mu) \rangle |\chi_\mu \rangle.
\label{eq:Tlmdef}
\eea

In order to derive an equation for
$T^D_{\lambda \mu}$, Eq.~(\ref{eq:M1}) is used to glean an
expression for $M^{(1)}$ in terms of the input to the equations:
$t_{\pi N}$, $t_{N N}$ and $f^{(1)}$. Using the facts about the
analytic structure of these amplitudes which were given above and
taking right and left residues of this expression as per
Eq.~(\ref{eq:Tlmdef}), it is found that:
\be
X_{\lambda \mu}^D=\langle \phi(\lambda)| \langle
\chi_\lambda| d_\lambda^{-1}
\gtb T_{\lambda \mu}^D \gtb
d_\mu^{-1} |\phi(\mu) \rangle |\chi_\mu \rangle,
\ee
with:
\bea
T_{\lambda \mu}^D &=&U_{\lambda \mu}^{(2)} + \left[ \sum_j
(1-v^X(\lambda) d_\lambda^{-1} \gtb) \bd{\lambda}{j} \fai{1}(j)
d_j^{-1} \right.\nn\\*
&+& \left. \sum_{\delta j} U_{\lambda
\delta}^{(2)} \gtb \tit{1}(\delta) d_{\delta}^{-1} d_{\bar{j}}
d_\pi \bd{\delta}{j} \fai{1}(j) \right] d_1 d_2 \nn\\*
& &(T_{NN}^D d_1 d_2  + 1) \left[\sum_{i \gamma}
\fir{1}(i) \bd{i}{\gamma} d_{\bar{i}} d_\pi \tait{1}(\gamma)
d_\gamma^{-1} \gtb U_{\gamma \mu}^{(2)} \right. \nn\\* &+&
\left. \sum_i
\fir{1}(i) d_i^{-1} \bd{i}{\mu} (1-\gtb v^X(\mu) d_\mu^{-1})
\right].
\label{eq:Tlm1}
\eea

We may also define $X_{N \mu}^D$ as the two-fragment matrix element
for transition from the two-body state in
which particle $\mu$ is a spectator to the state containing two
(distinguishable) nucleons:
\be
X_{N \mu}^D=(-i) \langle \psi_D| F_D^{(1)} d_\mu^{-1} \gtb
\mbox{Res}_{\mu \mbox{ pole}} \gtb d_\mu^{-1} |\phi(\mu) \rangle
|\chi_\mu \rangle d_\mu^{-1}. \ee
Once more it is found that:
\be
X_{N \mu}^D=\langle \psi_D| T_{N \mu}^D \gtb d_\mu^{-1}
|\phi(\mu) \rangle |\chi_\mu \rangle,
\ee
this time with:
\bea
T_{N \mu}^D&=&(T_{NN}^D d_1 d_2 + 1) \left[\sum_{j \alpha}
\fir{1}(j) \bd{j}{\alpha} d_{\bar{j}} d_\pi \tait{1}(\alpha)
d_\alpha^{-1} \gtb U_{\alpha \mu}^{(2)} \right. \nn\\
&+& \left. \sum_j \fir{1}(j)
d_j^{-1} \bd{j}{\mu} (1-\gtb v^X(\mu) d_\mu^{-1}) \right].
\label{eq:TNm}
\eea

Using the covariant AGS equations and Eq.~(\ref{eq:TNm}),
Eq.~(\ref{eq:Tlm1})  may now be simplified to:
\bea
T_{\lambda \mu}^D &=&\bd{\lambda}{\mu} d_1^{-1} d_2 ^{-1}
d_\pi^{-1} + \sum_j  (1-v^X(\lambda) d_\lambda^{-1} \gtb)
\bd{\lambda}{j} \fai{1}(j) d_j^{-1} d_1 d_2 T_{N \mu}^D \nn\\
&-& \sum_{\alpha j} \bd{\lambda}{\alpha} v^X(\alpha) d_\alpha^{-1}
d_{\bar{j}} d_\pi
\bd{\alpha}{j} \fai{1}(j) d_1 d_2 T_{N \mu}^D + \sum_\alpha
\bd{\lambda}{\alpha} t^{(1)}(\alpha) d_\alpha^{-1} \gtb T_{\alpha
\mu}^D .
\label{eq:Tlm}
\eea

Note that once again, this equation is equivalent to that
derived by AM \cite{AM83} and AB \cite{AB80}, but with terms
subtracted in order to eliminate the double-counting.

Finally, an equation for $T_{N \mu}^D$ is needed in
order to close the set of equations for the $NN-\pi NN$
system. Beginning with Eq.~(\ref{eq:TNm}) for $T_{N \mu}^D$, we
use the covariant AGS equations and Eq.~(\ref{eq:TNN}) for
$T_{NN}^D$ in order to obtain:
\bea
T_{N \mu}^D&=&\sum_j \fir{1}(j) d_j^{-1} \bd{j}{\mu} (1-\gtb
v^X(\mu) d_\mu^{-1}) + \bar{V} d_1 d_2 T_{N \mu}^D
\nn\\   &+& \sum_{j \alpha} \fir{1}(j) \bd{j}{\alpha}
d_{\bar{j}} d_\pi \tait{1}(\alpha) d_{\alpha}^{-1}
\gtb T_{\alpha
\mu}^D,
\eea
with $\bar{V}$ given by Eq.~(\ref{eq:Vbar}). Note the
changes in this equation as compared with AM's \cite{AM83} and
AB's \cite{AB80} result:
\begin{enumerate}
\item The $t^{(1)}(\alpha)$ appearing before $\fir{1}(j)$ has
been replaced by a $\tait{1} (\alpha)$.

\item  $V_{OPE}$ has been replaced by $\bar{V}$.

\item The term $\sum_j \fir{1}(j) \bd{j}{\mu} d_{\bar{j}} d_\pi
v^X(\mu) d_\mu^{-1}$ has been subtracted from the driving term.
\end{enumerate}

\section {Anti-symmetrization of the coupled equations}

\label {sec-antisymm}

At this stage the amplitudes $T_{NN}^D$, $T_{\lambda N}^D$, $T_{N
\mu}^D$ and $T_{\lambda \mu}^D$ are unphysical, since they have
not been anti-symmetrized. In this section we derive equations
for the anti-symmetrized amplitudes. These anti-symmetrized
equations apply to identical nucleons.

In order to anti-symmetrize the equations we must first derive
the correct anti-symmetrization procedure. It is by no means
self-evident that the ordinary non-relativistic procedure of
taking appropriate linear combinations of unsymmetrized
amplitudes will still be valid for these amplitudes which are
fully relativistic and obey four-dimensional integral equations.
The only way to discover the correct anti-symmetrization
procedure is to examine the Feynman Rules for the amplitudes.
Examination of the Feynman Rules for a two-fermion Green's
function shows that if
\be
G_{NN}^D(p_1',p_2';p_1,p_2)
\ee
is the Green's function which is obtained when two nucleons are
treated as distinguishable particles then:
\be
G_{NN}(p_1',p_2';p_1,p_2)=G_{NN}^D(p_1',p_2';p_1,p_2)-G_{NN}^D(p_2',p_1';p_1,p_2).
\ee

Applying LSZ reduction to each side of this equation, in order
to obtain an equation connecting the amplitudes involved,
then gives:
\be
T_{NN}=(1-P_{12}) T_{NN}^D,
\label{eq:TNNAS}
\ee
where $T_{NN}^D$ is the amplitude for distinguishable
particles and $P_{12}$ acting on any state interchanges the roles
of particles one and two in that state. Using the fact that:
\be (1-P_{12})^2=2(1-P_{12}),
\ee and that:
\be [P_{12},T_{NN}^D]=0,
\ee
Eq.~(\ref{eq:TNNAS}) gives:
\be
T_{NN}=\frac{1}{2} (1-P_{12}) T_{NN}^D (1-P_{12}).
\label{eq:TNNind}
\ee

A similar argument for the $NN \rightarrow NN \pi$
and $NN \pi \rightarrow NN \pi$ Green's functions shows:
\bea
\bfai{1}&=&\frac{1}{2} (1-P_{12}) {F^{(1)}_D}^\dagger
(1-P_{12})\\
M^{(1)}&=&\frac{1}{2} (1-P_{12}) {M^{(1)}_D} (1-P_{12}).
\label{eq:M1ind}
\eea

Consequently, if we now write:
\be
X_{\Delta N}^{\mbox {physical}} \equiv \langle \phi(2)| \langle
\chi_2| d_1 d_\pi \mbox{Res}_{(N_1 \pi)  \mbox{
pole}} d_1 d_\pi \bfai{1} |\psi_D \rangle,
\label{eq:XDNdef}
\ee
we find that:
\be
X_{\Delta N}^{\mbox {physical}}=\langle \phi(2)| \langle
\chi_2| d_1 d_\pi T_{\Delta N} |\psi \rangle,
\label{eq:XDelN}
\ee
where:
\be
T_{\Delta N}=\frac{1}{\sqrt{2}} (T_{2 N}^D-P_{12} T_{1 N}^D),
\ee
and $|\psi \rangle$ is a fully anti-symmetrized $NN$
wave-function:
\be
|\psi \rangle=\frac{1}{\sqrt{2}} (1-P_{12}) |\psi_D \rangle.
\ee
(Note that the choice of final-state spectator in
Eq.~(\ref{eq:XDNdef}) makes no difference to the result, beyond an
overall minus sign.)
Similarly:
\bea
X_{\Delta d}^{\mbox {physical}} &\equiv& \langle \phi(2)| \langle
\chi_2| d_1 d_\pi \mbox{Res}_{(N_1 \pi) \mbox{ pole}} d_1 d_\pi
M^{(1)} d_1 d_2 \mbox{Res}_{(N_1 N_2) \mbox{ pole}} d_1 d_2
|\phi(3)_D \rangle |\chi_\pi \rangle\\
&=& \langle \phi(2)| \langle
\chi_2| d_1 d_\pi  T_{\Delta d} d_1 d_2
|\phi(d) \rangle |\chi_\pi \rangle,
\eea
where:
\bea
T_{\Delta d}&=&\frac{1}{\sqrt{2}} (T_{2 3}^D-P_{12}
T_{1 3}^D),\\
|\phi(d) \rangle&=&\frac{1}{\sqrt{2}} (1-P_{12}) |\phi(3)_D
\rangle.
\eea

We now take the definitions of
the distinguishable particle matrix elements
$X_{22}$, $X_{N2}$, $X_{32}$, $X_{NN}$, $X_{33}$, $X_{3N}$ and
$X_{N3}$, replacing everywhere the distinguishable particle
amplitudes with their identical-particle counterparts
(\ref{eq:TNNind})--(\ref{eq:M1ind}). This yields definitions for
the identical-particle matrix elements  $X_{\Delta
\Delta}^{\mbox{physical}}$, $X_{N \Delta}^{\mbox{physical}}$,
$X_{d \Delta}^{\mbox{physical}}$, $X_{NN}^{\mbox{physical}}$,
$X_{d d}^{\mbox{physical}}$, $X_{d N}^{\mbox{physical}}$ and $X_{N
d}^{\mbox{physical}}$. When taken together with the definitions
$X_{\Delta N}^{\mbox{physical}}$ and $X_{\Delta
d}^{\mbox{physical}}$ given above this produces nine definitions
which may be written in the form:
\be
X_{AB}^{\mbox{physical}}=\langle \psi_A|T_{AB}|\psi_B \rangle,
\ee
where $A,B=N,\Delta,d$,
\bea
|\psi_N \rangle&=&|\psi_{NN} \rangle\\
|\psi_\Delta \rangle&=&d_1 d_\pi |\phi(2) \rangle |\chi_2
\rangle\\
|\psi_d \rangle&=&d_1 d_2 |\phi(d) \rangle |\chi_\pi \rangle;
\eea
and:
\bea
T_{NN}=T_{NN}^D; \quad T_{Nd}=T_{N 3}^D; &\quad& T_{d N}=T_{3
N}^D;
\quad T_{dd}=T_{33}^D;\\
T_{\Delta N}=\frac{1}{\sqrt{2}}(T_{2 N}^D-P_{12} T_{1 N}^D);
&\quad& T_{N \Delta}=\frac{1}{\sqrt{2}}(T_{N 2}^D-T_{N 1}^D
P_{12});\\ T_{\Delta d}=\frac{1}{\sqrt{2}}(T_{2 3}^D-P_{12} T_{1
3}^D); &\quad& T_{d \Delta}=\frac{1}{\sqrt{2}}(T_{3 2}^D-T_{3
1}^D P_{12});\\ T_{\Delta \Delta}&=&T_{2 2}^D-T_{2 1}^D P_{12}.
\eea

Equations for these amplitudes may then be derived from the
Eqs.~(\ref{eq:TlN}), (\ref{eq:TNN}), (\ref{eq:TNm}) and (\ref{eq:Tlm})
for $T_{\lambda N}^D$, $T_{NN}^D$, $T_{N \mu}^D$ and $T_{\lambda
\mu}^D$. We find that the set of coupled equations:
\bea
T_{\Delta N}&=&\sqrt{2} \left[f^\dagger (1) d_1^{-1} - Y \right]
(1+d_1 d_2 T_{NN}) - P_{12} t(2) d_1 d_\pi T_{\Delta N} + \sqrt{2}
t(3) d_1 d_2 T_{d N},\label{eq:ceq1}\\
T_{d N}&=&\left[ \sum_k f^\dagger (k)
d_k^{-1} - Y \right] (1 + d_1 d_2 T_{NN}) + \sqrt{2} t(2)
d_1 d_\pi T_{\Delta N},\\
T_{N N}&=&\bar{V} (1 + d_1 d_2 T_{NN}) + \sqrt{2}
f(1) d_2 d_\pi \tilde{t}^{\dagger}(2) d_1 d_\pi T_{\Delta
N} + \sum_k f(k) d_{\bar{k}} d_\pi t(3) d_1 d_2 T_{d N},
\eea
describe the coupled $NN \rightarrow \pi NN$ and $NN
\rightarrow NN$ two-fragment processes. Meanwhile the
equations:
\bea
T_{\Delta \Delta}&=&-P_{12} d_1^{-1} d_2^{-1} d_\pi^{-1}
+  \sqrt{2} \left[
f^\dagger(1) d_1^{-1} - Y \right] d_1 d_2 T_{N \Delta}\nn\\
&& - P_{12} t(2) d_1 d_\pi T_{\Delta \Delta} + \sqrt{2} t(3) d_1 d_2
T_{d \Delta}\\
T_{d \Delta}&=&\sqrt{2} d_1^{-1} d_2^{-1} d_\pi^{-1}
+ \left[\sum_k f^\dagger(k) d_k^{-1} - Y \right] d_1 d_2 T_{N
\Delta} + \sqrt{2} t(2) d_1 d_\pi T_{\Delta \Delta}\\
T_{N \Delta}&=&\sqrt{2}
f(1) d_1^{-1} (1 - d_1 d_\pi v^X(2)) + \bar{V} d_1 d_2 T_{N
\Delta}
+ \sqrt{2} f(1) d_2 d_\pi \tilde{t}^\dagger (2) d_1 d_\pi
T_{\Delta \Delta} \nn\\
&& + \sum_k f(k) d_{\bar{k}} d_\pi t(3) d_1
d_2 T_{d \Delta}\\
T_{\Delta d}&=&\sqrt{2} d_1^{-1} d_2^{-1} d_\pi^{-1} + \sqrt{2}
\left[ f^\dagger(1) d_1^{-1} - Y \right] d_1 d_2 T_{N d} - P_{12}
t(2) d_1 d_\pi T_{\Delta d}\nn\\
&&  + \sqrt{2} t(3) d_1 d_2 T_{d
d}\\
T_{d d}&=&\left[\sum_k f^\dagger (k)
d_k^{-1} - Y \right] d_1 d_2  T_{N d} + \sqrt{2} t(2) d_1 d_\pi
T_{\Delta d}\\
T_{N d}&=&\sum_k f(k) d_k^{-1} + \bar{V} d_1 d_2 T_{N d} +
\sqrt{2} f(1) d_2 d_\pi \tilde{t}^\dagger(2) d_1 d_\pi T_{\Delta
d}\nn\\
&&+ \sum_k f(k) d_{\bar{k}} d_\pi t(3) d_1 d_2 T_{d d}, \label{eq:ceqf}
\eea
describe the $NN \pi \rightarrow NN
\pi$ and $NN \rightarrow NN \pi$ two-fragment processes. Here:
\be
Y=\sum_k v^X(\bar{k}) d_\pi f^\dagger(k).
\ee
Note that the t-matrices and vertices
used as input to these equations are always the one-particle
irreducible amplitudes, $t^{(1)}$, $f^{(1)}$ and $\fai{1}$. An
apparent exception is the inclusion of the modified one-particle
irreducible t-matrix $\tait{1}$, which is used in some places in
the equations. But, $\tait{1}$ is, of course, defined in terms of
$t^{(1)}$ and $f^{(1)}$ by Eqs.~(\ref{eq:tildet2}) and
(\ref{eq:vR}).

\section {Specification of input amplitudes and implicit inclusion of
other hadronic states}

\label{sec-inpspec}

Equations (\ref{eq:ceq1})--(\ref{eq:ceqf}) are a set of two-fragment
scattering equations for the processes:
\begin{equation}
\left. \begin{array}{rr} N + N + \pi\\ N  + N
\end{array} \right \}
 \longrightarrow
\left \{
\begin{array}{ll} N + N + \pi\\  N  + N
\end{array} \right.
\label{eq:reac}
\end{equation}
Each Feynamn diagram in which no more than one explicit pion appears
is included once and only once in the equations. Hence these four-dimensional
$NN-\pi NN$ equations obey two and three-body unitarity and contain no
double-counting, thus providing a framework in which predictions for scattering
observables in the $NN-\pi NN$ system can be made.

The first step in making such predictions is the specification of the input
amplitudes $t_{\pi N}^{(1)}$ and $\fir{1}$. In Appendix~\ref{ap-input} the
one-explicit-pion content of $t_{\pi N}^{(1)}$ and $\fir{1}$ is elucidated  by
the Taylor method, demonstrating that the ampitudes which need to be specified
are, in fact, $t_{\pi N}^{(2)}$ and $\fir{2}$. Now, $t_{\pi N}^{(2)}$ and
$\fir{2}$ could have their {\em two}-explicit-pion content analyzed (see
e.g.,\cite{AP86} for the $\pi N$ case in time-ordered perturbation theory).
However, even were we able to solve the resulting set of coupled $\pi N-\pi \pi
N$ equations together with Equations (\ref{eq:ceq1})--(\ref{eq:ceqf}) the
results would still only include the effects of pion exchange. It is well known
that other effects, e.g.~heavy-meson exchange and delta degrees of freedom, are
important in nuclear forces, and it is not clear how such mechanisms would be
included in this description.

Therefore  in order to implicitly include some of this additional physics we
advocate calculating $\fir{2}$ and $\tpni{2}$ in some other (hadronic or
quark) model and/or parametrizing them in such a way as to simplify the
solution
of Eqs. \@(\ref{eq:ceq1})--(\ref{eq:ceqf}). Note that since, by definition,
these amplitudes have no one-explicit-pion content, such an approach does not
affect $NN$ or $\pi NN$ unitarity. Nor, if it is implemented correctly, should
it introduce any double-counting.

Firstly, the bare $\pi NN$ form factor $\fir{2}$ may be taken from some model
of QCD, as long as care is taken with any pionic corrections so that
overcounting is avoided. Alternatively, $\fir{2}$ may simply be parametrized.

Secondly, two approaches may be taken with $\tpni{2}$. On the one hand, a
separable potential may be postulated for $\tpni{2}$ and the free parameters of
the separable form fitted to the $\pi N$ scattering data. (Note that it may be
necessary to use $\pi N$ scattering data in the $P_{11}$ channel in order to
fix
any parameters in $\fir{2}$.)  In this way known mechanisms of the $\pi-N$
interaction, such as $\rho$ and $\omega$ exchange and the $\Delta$ resonance,
are implicitly included insofar as they contribute to the actual data. On the
other hand, if we wish to know {\em exactly} what physics is included in the
input $\pi N$ interaction a meson exchange model of $\pi N$ scattering, with
the
$\Delta$ resonance built in, may be constructed (see e.g.~\cite{JP91}) and a
separable expansion of the resulting amplitude made, via the
Ernst-Shakin-Thaler
technique \cite{Er73A,Er73B}. Such an expansion has already been accomplished
for the three-dimensional Paris $NN$ potential~\cite{HP84}. In the $\pi N$ case
Pearce and Afnan have used the extension of Ernst, Shakin and Thaler's
technique developed by Pearce~\cite{Pe87} in order to obtain a separable
expansion of the $\pi N$ interaction in the Cloudy Bag Model~\cite{AP89}.
The work of Rupp and Tjon~\cite{TR88} shows that making a separable
expansion of
a Bethe-Salpeter amplitude is not computationally intensive, thus once a
meson-exchange model of the $\pi N$ system has been used to calculate
scattering
via the Bethe-Salpeter Equation a separable expansion of the resulting
amplitude
should not be a difficult task.

Thirdly, so far we have regarded $\tnni{1}$ as part of the solution to
Eqs.~(\ref{eq:ceq1})--(\ref{eq:ceqf}). However, this amplitude is also needed
 in
order to construct the kernel of these equations.  Indeed, it is $\tnni{1}$s
appearance in the kernel (as $t_{NN}^{(1)}$) which ensures that higher pionic
processes, such as crossed two-pion exchange, are included in the solution to
the coupled equations (see Figure \ref{fig-XedTPE}). In theory it is possible
to
``bootstrap" this theory up and so generate an exact solution to the non-linear
equation for $\tnni{1}$. However even if this could be achieved, important
physics would be missing from the resulting $NN$ interaction. In order to
simultaneously include some of these missing mechanisms and snap the
``bootstrap", we advocate the same approach to the input
$t_{NN}^{(1)}$ as to $\tpni{1}$. Either Eq.~(\ref{eq:BSNN1}) for $\tnni{1}$
should be used and a separable potential for $\tnni{2}$ fitted to the
experimental data, or a four-dimensional covariant separable expansion of some
meson-exchange $NN$ amplitude should be made. As in the $\pi N$ case, since the
input $t_{NN}^{(1)}$ occurs only in the presence of a spectator pion such an
approach will not destroy $NN$ or  $NN \pi$ unitarity, nor will it, if
carefully
implemented, introduce double-counting. The consistency of such a procedure can
be checked by comparing the solution $\tnni{1}$ with the input $t_{NN}^{(1)}$.

If the amplitudes $t_{NN}^{(1)}$, $\tpni{1}$ and $\fir{1}$ are constructed in
this way then (provided a separable expansion for $v^X$ is also made)
Eqs.~(\ref{eq:ceq1})--(\ref{eq:ceqf}) become a set of coupled Bethe-Salpeter
equations for the $NN-\pi NN$ system. This is completely analagous to the
situation in the non-relativistic three-body Faddeev equations \cite{Fa61A}. By
using separable expansions for the input amplitudes these may be reduced to the
Lovelace equations \cite{Lo64}, which are a set of coupled two-body equations.

One question which must be answered is which energy domain any
separable expansion of
$t_{NN}^{(1)}$ and $t_{\pi N}^{(1)}$ should be made in. Suppose that Figure
\ref{fig-FDpiece} is a piece of a Feynman Diagram summed in the
four-dimensional $NN-\pi NN$ equations, with $\alpha$ the spectator particle
label ($\alpha=\pi,N$) and $k_\alpha$ its four-momentum. The analytic
expression for this Feynman diagram would therefore contain a piece:
\be
\int d^4 k_\alpha \cdots G_{\bar{\alpha}} (P-k_\alpha) \, t^{(1)}_\alpha
(s_\alpha) \, d_\alpha (k_\alpha)
\, G_{\bar{\alpha}} (P-k_\alpha) \cdots
\label{eq:FDpiece}
\ee
where
\bea
P&=&(\sqrt{s},\vec{0})\\
s_\alpha=(P-k_\alpha)^2&=&(\sqrt{s} - k_\alpha^0)^2 - k_\alpha^2
\eea
is the total four-momentum in the three-body centre of mass, and
$G_{\bar{\alpha}} (P-k_\alpha)$ is the product of the two propagators of the
interacting particles.

Since $k_\alpha^0 \in (-\infty,\infty)$ the two-body energy
which is the argument of $t^{(1)}_\alpha$ ranges from $-\infty$ to
$\infty$. However, suppose that a separable expansion for $t^{(1)}_\alpha$ has
been made, i.e.:
\be
t^{(1)}_\alpha (s_\alpha)=\sum_i g_i^\dagger \frac{1}{s_\alpha^+ - m_i^2} g_i,
\ee
with $g_i$ an appropriate form factor implicitly dependent on the
relative four-momentum of the two particles in the interacting subsystem.
In order to perform the $k_\alpha^0$ integration we examine the analytic
structure of the integrand in the complex $k_\alpha^0$ plane.

Apart from the dependence listed in Eq.~(\ref{eq:FDpiece}), the only place
$k_\alpha$ appears in the integrand is via its presence in the form factors
which occur immediately preceding and following the piece of diagram which is
explicitly written in (\ref{eq:FDpiece}). Unless these are $\pi NN$ form
factors
their analytic structure in the relative momentum variable must give no
contribution to unitarity, and therefore their $k_\alpha$-dependence may be
ignored. If one (or both) form factors is a $\pi NN$ form factor then there
is a $\pi N$ cut (or cuts) to be added to the analytic structure in the lower
half of the $k_\alpha^0$ plane that is listed below. However, the existence of
such cuts makes no difference to the final conclusion obtained here.
Consequently, for the purposes of this argument, we ignore any
$k_\alpha$-dependence of the form factors.

The poles of the integrand in the lower half-plane are at:
\bea
\sqrt{m_\alpha^2 + |\vec{k}_\alpha|^2} - i\epsilon \label{eq:pole}\\
\sqrt{s} + \sqrt{m_i^2 + |\vec{k}_\alpha|^2} - i\epsilon
\eea
while the two-body propagator contributes a cut beginning at:
\be
\sqrt{s} + m_{\bar{\alpha}} - i\epsilon
\ee
where $m_{\bar{\alpha}}$ is the mass of the two-body system. Provided that
$s$ is in the region for particle-particle scattering the piece of
analytic structure giving the most significant contribution to the integral is
the pole (\ref{eq:pole}).

Thus the important two-body energies $s_\alpha$ are those satisfying
$\sqrt{s_\alpha} < \sqrt{s} - m_\alpha$, in agreement with the case of
non-relativistic few-body physics~\cite{AB85}. Therefore if $\sqrt{s}$ is
restricted to:
\be
2m \leq \sqrt{s} \leq 2m + 2\mu
\ee
then $\pi N$ and $NN$ amplitude parametrizations based on information in
the region below the respective pion production thresholds should give
accurate results when used in the solution of
Eqs.~(\ref{eq:ceq1})--(\ref{eq:ceqf}).

Finally, it is clear that $NN$ interactions mediated by meson exchanges which
are totally non-pionic are not included in our equations. These exchanges are
known to be important effects in the medium to short range part of the
nucleon-nucleon interaction. They could be included by extending the arguments
given here for the pion in order to expose one-explicit-meson states for {\em
all} mesons of importance. However, at this stage such an approach would be
unnecessarily thorough, since we do not wish to make predictions for the
reactions:
\bea
NN + X \leftrightarrow NN\\
X + d \rightarrow X + d,
\eea
where $X$ is any meson known to be important in one-boson-exchange $NN$
potentials! Therefore we can afford to merely add a phenomenological
heavy-meson exchange piece to the $\tnni{2}$ given by Eq.(\ref{eq:finalT2}), as
was done in the three-dimensional $NN-\pi NN$ case by Avishai and Mizutani
\cite{AM83,La87}. Note that some care must be taken in making this addition.
For
instance, the piece of $\rho$ exchange arising from the exchange of two
uncorrelated pions is separately included in the $NN-\pi NN$ equations,
therefore if a $\rho$ is phenomenologically added to our formalism it will not
necessarily have the same parameters as in $NN$ OBE potentials.

\section {Conclusion}

\label {sec-conclusion}

In this paper we have derived coupled four-dimensional covariant scattering
equations for the $NN-\pi NN$ system, using the modified Taylor method of
classification of diagrams \cite{AP94}.  These equations sum all the covariant
perturbation theory diagrams which include one explicit pion once and only
once.
They therefore:
\begin{enumerate}
\item Are four-dimensional integral equations.

\item Are covariant, not only on-shell,
but also have off-shell covariance in the manner dictated by the Feynman
graphical expansion.

\item Are completely free of the double-counting problems of some
previous four-dimensional $NN-\pi NN$ equations.

\item Obey $NN$ and $NN \pi$ unitarity, by explicit construction.
\end{enumerate}

The double-counting subtractions found to be necessary are a consequence of
attempting to derive a set of coupled equations which describe, in a unified
way, the reactions:
\begin{eqnarray}
N + N &\rightarrow& N + N\nn\\
N + N &\leftrightarrow& N + N + \pi \nn\\
N + N + \pi &\rightarrow& N + N + \pi,
\label{eq:processes}
\end{eqnarray}
by summing all covariant perturbation theory Feynman diagrams for the
processes (\ref{eq:processes}) that involve one explicit pion or no explicit
pions. Were one content merely to take the Bethe-Salpeter equation for $NN$
scattering and choose a set of Feynman diagrams for the kernel $\tnni{2}$ no
double-counting would arise. However, such an approach does not satisfy
three-body unitarity, since it excludes certain $s$-channel two-particle
irreducible diagrams which contain one explicit pion. Only by deriving a set of
coupled equations which sum {\em all} covariant perturbation theory diagrams
containing zero or one explicit pion(s), as we have done here, will three-body
unitarity be obeyed.

The modified Taylor method used to make this derivation is completely
general and
so could also be used to obtain equations for other few-hadron processes. For
example, the $\pi N-\pi \pi N$ problem and pion photoproduction would
both be amenable to a derivation by this method.

The double-counting exposed and eliminated here occurred in four-dimensional
$NN-\pi NN$ equations. The implication of this double-counting for the
``standard" three-dimensional $NN-\pi NN$ equations is not entirely clear. On
the one hand, an obvious way to obtain three-dimensional equations from the
ones given here is to take the two and three-particle Green's functions
appearing in these equations and replace them by three-dimensional Green's
functions, as suggested by, for example, Blankenbecler and Sugar \cite{BbS66}.
It is apparent that the equations thus produced will be different to the
standard $NN-\pi NN$ equations, since they will still include subtractions for
double-counting. On the other hand, the standard $NN-\pi NN$
equations may also be derived by classifying diagrams in time-ordered
perturbation theory, as was mentioned in the Introduction. If this approach
is taken the resulting equations do not double-count any time-ordered
perturbation theory diagrams. Hence, one may question whether the subtractions
which would appear were one to apply a three-dimensional reduction to our
$NN-\pi NN$ equations are really necessary.

Subject to this issue of the relation to a three-dimensional calculation, our
results may have ramifications beyond few-hadron processes. For instance,
the double-counting present in the unsubtracted four-dimensional
$NN-\pi NN$ equations may have implications for pion absorption and scattering
on larger nuclei, as has been discussed by, for example Kowalski et
al.\cite{Ko79}. Therefore the remedies for this double-counting discussed
above may well prove applicable in these larger systems.

The set of equations obtained here are (apart from the minor point
discussed in Section~\ref{sec-c1calc}) equivalent to those previously
obtained by Kvinikhidze and Blankleider (KB) \cite{BK94A,BK94B}, although we
have anchored our derivation more firmly in Taylor's original work.
Furthermore, the final set of coupled equations given here are ready for
computation, being written in the form of two-fragment scattering equations.

These scattering equations require as their input only the $\pi-N$
t-matrix and dressed $\pi NN$ vertex. However, since the $NN$ t-matrix
is present in the equations both within the kernel and as part of the solution
it appears that some way of ``bootstrapping" the theory up must be found.
In Section~\ref{sec-inpspec} we discussed how this bootstrap might be avoided
and came to the inclusion that making a separable expansion of a model $NN$
amplitude at energies up to the pion-production threshold should provide a good
input $NN$ amplitude. Even once the bootstrap problem is resolved in this way
the
computation is still somewhat formidable, involving, as it does, the numerical
solution of a set of coupled four-dimensional integral equations. Nevertheless,
the equations derived here (together with those obtained by KB) represent the
first complete and correct summation of the one-explicit-pion sector of the
theory of the $NN-\pi NN$ system. Not until these equations are solved will it
be clear how much of the physics of the $NN-\pi NN$ system is attributable
to the one-explicit-pion sector. Therefore, we believe that the numerical work
involved here, although at first sight somewhat daunting, is necessary.

Despite the fact that we have had to expose some $NN \pi \pi$ states to
remove double-counting on the $NN \pi$ level we have, in general,
refrained from entering the two-pion sector of the theory. This has meant
that the amplitudes $M^{(3)}$, $\bfai{3}$ and $T^{(3)}$ have been completely
ignored. The modification to the above equations to include such ``three-body"
mechanisms is one obvious way in which these equations could be extended.

We note that mesons other than the pion and baryonic states such as the
$\Delta$ may be implicitly included in the equations via their (possibly
parametrized) presence in input amplitudes. They may be included explicitly if
the Lagrangian is extended to one involving all hadrons of interest. Such a
modification of the underlying field theory merely results in a proliferation
of possible diagrams, rather than any fundamental change in the way the
derivation proceeds.

However, such improvements are in the future. Until the equations as
they stand here are solved, and it is seen how well they reproduce the
experimental data in the $NN-\pi NN$ system, we shall not know which, if
any, of these more sophisticated mechanisms are necessary for a correct
description of the $NN-\pi NN$ system dynamics. The next stage of this
work must therefore be an effort to solve the equations given here
numerically.

As a first step towards this goal we have recently used the formalism
developed here to perform a calculation of $\phi \phi$ scattering in a
scalar $\phi^2 \sigma$ field theory. Numerical methods and results will be
detailed in a later paper \cite{AP96}.

\appendix

\section {Propagator dressing}

\label {ap-dress}

In this appendix we discuss how to accomplish the dressing of all the particles
involved. The procedure used is similar to that developed by Afnan and
Blankleider
\cite{AB80,AB85}.

Consider the amplitude $\At{m}{n}$ for the $m
\rightarrow n$ transition. The diagrams contributing to
$A$ may be of two types:
\begin{enumerate}
\item Those for which a self-energy contribution on one leg, which we consider
to be leg $i$ in the initial state, may be isolated. (The argument is exactly
the same for self-energy contributions on legs in the final state.) These
diagrams are of the general form: \be
\At{m}{n} d_i^{(0)} \Sigma_i, \ee where $d_i^{(0)}$ and $\Sigma_i$ are the
undressed propagator and self-energy of particle $i$. (If there are $j$
nucleons in the initial state then for $i \leq j$ $d_i^{(0)}$ and
$\Sigma_i$ represent a nucleon propagator and self-energy, otherwise they
represent a pion propagator and self-energy.) See Figure
\ref{fig-Selfenergy} for an example of such a diagram.

\item The diagrams which cannot have a self-energy contribution isolated in
this way. We denote these by: \be {\At{m}{n}}_{(\tilde{i})}  \ee
\end{enumerate}

Therefore,
\bea
\At{m}{n}=\At{m}{n} d_i^{(0)}
\Sigma_i + {\At{m}{n}}_{(\tilde{i})},\\
\implies
\At{m}{n} d_i^{(0)}={\At{m}{n}}_{(\tilde{i})} d_i. \eea  Since,
\be d_i=d_i^{(0)} (1-\Sigma_i d_i^{(0)})^{-1}=({d_i^{(0)}}^{-1} -
\Sigma_i)^{-1}=Z_i d_i^R,
\ee where $Z_i$ is the wave function renormalization for particle $i$, and
$d_i^R$ is the renormalized propagator with unit residue at the pole of
$d_i$ which corresponds to the physical mass of particle $i$.

Repeating this procedure for all external legs leads to the following result:
\be  G_n
\At{\tilde{m}}{\tilde{n}} G_m=G_n^{(0)}
\At{m}{n} G_m^{(0)},
\ee  where:
\bea G_k&=&\prod_{i=1}^k d_i Z_i^{-\half},\\ G_k^{(0)}&=&
\prod_{i=1}^k d_i^{(0)},
\eea
and $\At{\tilde{m}}{\tilde{n}}$ has had all the external bubbles removed
from it, and factors of
$Z^{\half}$ included in it, in order to make it agree with Eq.~(\ref{eq:LSZ}).

This result allows us to consider all initial and final state legs as
fully-dressed, provided that we work with the amplitude
$\At{\tilde{m}}{\tilde{n}}$ which has no bubbles on these initial or final
state legs. Bubbles may also be eliminated from the internal legs of
$\At{\tilde{m}}{\tilde{n}}$ by a similar procedure, as follows. Consider
any set of diagrams contributing to
$\At{\tilde{m}}{\tilde{n}}$ which are
$(k-1)$-particle irreducible but admit an internal
$k$-particle cut. According to the last internal cut lemma these diagrams
may be expressed uniquely as:
\be
\At{k}{\tilde{n}} G_{k}^{(0)} \At{\tilde{m}}{\tilde{k}}.
\ee
Using the same arguments as above any bubbles appearing in the initial state
of $\At{k}{\tilde{n}}$ may be amputated and placed in $G_k$. Consequently, we
obtain for this sum:
\be
\At{\tilde{k}}{\tilde{n}} G_{k}
\At{\tilde{m}}{\tilde{k}}.
\ee

Therefore we may always work with fully-dressed particles, provided that we
consider all amplitudes to be of the type $\At{\tilde{m}}{\tilde{n}}$. In this
paper we have adopted this approach, and so all amplitudes are of the
$\At{\tilde{m}}{\tilde{n}}$ type, even though the tildes are never displayed
explicitly.

\section {Input to the $NN-\pi NN$ equations: the $\pi NN$ vertex and the
$\pi-N$ amplitude.}

\label {ap-input}

In this appendix we consider the two amplitudes which are  required as input to
the $NN-\pi NN$ equations: the one-particle irreducible $\pi NN$ vertex,
$\fir{1}$, and the one-particle irreducible
$\pi-N$ t-matrix, $t_{\pi N}^{(1)}$.

Firstly, consider $\fir{1}$. Our aim is to apply the Taylor method to this
vertex in order to derive an integral equation for it. Since we are examining
the two-cut structure of $\fir{1}$, we have
$m=2$, $n=1$ and $r=2$. Consequently, since we are dealing with fully-dressed
particles all one-to-one amplitudes are zero and so classes $C_3$, $C_4$ and
$C_5$ are all empty. Therefore double-counting in classes $C_4$ and $C_5$
cannot
arise, even though $n \leq r$. The sum of class
$C_1$ in this case is clearly the two-particle irreducible
$\pi NN$ vertex
$f^{(2)}$. The sum of class $C_2$ is found, via the last internal cut lemma, to
be:
\be  f^{(2)} d_N d_\pi t_{\pi N}^{(1)}.
\ee  Therefore, Taylor's method gives the following equation for $\fir{1}$:
\be
\fir{1}=\fir{2}+\fir{2} d_N d_\pi t_{\pi N}^{(1)}.
\label {eq:vertex}
\ee

Taylor's method could now be used in order to extract the structure of
$f^{(2)}$, but we prefer to not consider three-body unitarity for this
amplitude. Instead the expression for $f^{(2)}$ will be extracted from the
Lagrangian under consideration.

However, we do wish to discuss the two-particle cut structure of the
1PI $\pi-N$ amplitude, $\tpni{1}$, using the Taylor method. Observe that
classes
$C_3$, $C_4$ and $C_5$ are again empty, since all particles are fully dressed.
Applying the Taylor method, it is clear that the sum of class $C_1$ is the 2PI
amplitude
$\tpni{2}$. Also, the LICL may be applied to class
$C_2$ in order to obtain:
\be
\tpni{2} d_N d_\pi
\tpni{1};
\ee  and therefore the following Bethe-Salpeter type equation for $\tpni{1}$ is
obtained:
\be
\tpni{1}=\tpni{2}+\tpni{2} d_N d_\pi \tpni{1}.
\label {eq:tnp}
\ee  As mentioned above in the case of $\fir{2}$, the structure of $\tpni{2}$
may be investigated using similar techniques as have been used for $\tpni{1}$.
On the other hand, if we are not concerned with $\pi \pi N$ unitarity in the
$\pi-N$ amplitude we may just extract some model-dependent result for
$\tpni{2}$ from the Lagrangian under consideration.

The amplitude $\tpni{1}$ is, however, only part of the full $\pi-N$ amplitude,
$\tpni{0}$. Applying Taylor's method to $\tpni{0}$ in the same fashion as
above, reveals that:
\be
\tpni{0}=\tpni{1}+\fai{1} d_N \fir{0}.
\ee  A brief examination of the zero-particle irreducible $\pi NN$ amplitude,
$\fir{0}$, shows that the only Taylor class of diagrams contributing to
$\fir{0}$
which is non-empty is $C_1$, which sums to give the 1PI
$\pi NN$ amplitude, $\fir{1}$. Therefore,
\be
\fir{0}=\fir{1}.
\label{eq:f0}
\ee  Consequently, the full $\pi-N$ amplitude $\tpni{0}$ is given by:
\be
\tpni{0}=\tpni{1}+\fai{1} d_N \fir{1},
\ee where $\tpni{1}$ is the non-pole part of the $\pi-N$ t-matrix, given by
Eq.~(\ref{eq:tnp}), and $\fai{1} d_N
\fir{1}$ is the pole part of the $\pi-N$ t-matrix.

\section {Details of the calculation of $\lowercase{c}_1 \cap C_5^{\{N2'\}}$}

\label {ap-calcdet1}

The portion of $\ttpni{2}$ which is 2PR in the $N'
\leftarrow \pi' + \pi + N$ channel may be written, using the reverse of the
LICL, the "first internal cut lemma", as:
\be
\fir{1}_{\pi_1} d_1 d_{\pi_1} P^{\dagger}
\label {eq:c11}
\ee   where $P^{\dagger}$ is an $N \pi_2
\rightarrow N \pi_1 \pi_2$ amplitude which is two-particle irreducible in the
$s$-channel and the channel:
$$\pi_1' + N'\leftarrow N + \pi_2 +\pi_2',$$ as well as being 1PI in the
channel: $$\pi_2' + N \leftarrow N' + \pi_2 +\pi_1'.$$
Consequently, we might expect the diagrams:
\be
\fir{1}_{\pi_2}(1) d_{\pi_2}
\fir{1}_{\pi_1}(2) d_1 d_{\pi_1} P^{\dagger}(2) d_2 d_{\pi_2}
\fsai{1}{\pi_2}(1)
\label{eq:appdc1}
\ee and:
\be
\fir{1}_{\pi_2}(1) d_{\pi_2}
\fir{1}_{\pi_1}(2) d_1 d_{\pi_1} P^{\dagger}(2) d_1 d_{\pi_2}
\tilde{t}^{(1)}_{\pi_2 N} (2) d_2 d_{\pi_2}
\fsai{1}{\pi_2}(1)
\label{eq:appdc2}
\ee
to be included in $c_1 \cap \cfi{N2'}$. Consider first (\ref{eq:appdc1}). If
this term is to be part of $\cfi{N2'}$ the pion absorbed on $N2'$ must be
``hidden" in another amplitude, since if the diagram is to contribute to
$\cfi{N2'}$ it must take the form:
\be
\fir{1}(2) d_1 d_\pi \bfai{2}.
\ee There are three ways of ``hiding" the pion:
\begin{enumerate}
\item
Place it in an $N-N$ t-matrix. Taking this approach shows that the diagram:
\be
\fir{1}_{\pi}(1) d_\pi \fir{1}_{\pi'}(2) d_1 t^{(1)}_{\pi N}(2) d_{\pi'} d_1
\fsai{1}{\pi'}(2) d_\pi d_2
\fsai{1}{\pi}(1)
\ee
is in $c_1 \cap \cfi{N2'}$, since it occurs in both $c_1$ and the term:
\be
\fir{1}(2) d_\pi d_1 \tnni{1} d_1 \fai{1}(2)
\ee
of $\cfi{N2'}$.

\item The second way of hiding the second pion is to place it in the $\pi-N$
t-matrix. That is, to note that the diagrams:
\bea
\fir{1}_{\pi_2}(1) d_{\pi_2} d_2 \fir{1}_{\pi_1}(2) d_{\pi_1} d_1
t^{(1)}_{\pi_1
\pi_2} d_{\pi_1}
\fsai{1}{\pi_1}(2) d_{\pi_2}
\fsai{1}{\pi_2}(1) \mbox { and }\nn\\
\fir{1}_{\pi_2}(1) d_{\pi_2} d_2 \fir{1}_{\pi_1}(2) d_{\pi_1} d_1
t^{(1)}_{\pi_1 \pi_2} d_{\pi_2}
\fsai{1}{\pi_1}(2) d_{\pi_1}
\fsai{1}{\pi_2}(1)
\eea is included in both $c_1$ and:
\be
\fir{1}(2) d_\pi d_2 \tpni{1}(1) d_\pi \fai{1}(2),
\ee   which is part of $\cfi{N2'}$.

\item Finally, one can "hide" the pion in a three-body force. This suggests
that if the term:
\be
\fir{1}(2) d_\pi d_1 M^{(3)}_{1} d_2 d_\pi \fai{1}(1)
\label{eq:c13BF}
\ee
is included in $\cfi{N2'}$ then a number of diagrams in $c_1$ will also
occur in $\cfi{N2'}$. Precisely which diagrams from
$c_1$ are included in (\ref{eq:c13BF}) though? The additional irreducibility
constraints placed on $M_1^{(3)}$ mean that only that part of $c_1$,
$\tilde{c}_1$, which is given by:
\be
\tilde{c}_1=\fir{1}_{\pi_2}(1) d_{\pi_2} d_2 \fir{1}(2) d_1 d_{\pi_1}
\tilde{P}^\dagger(2) d_{\pi_2}
\fsai{1}{\pi_2}(1),
\ee where $\tilde{P}^{\dagger}$ is 1PI in both the channels:
$$N' + N \leftarrow \pi_1' + \pi_2' + \pi_2 \mbox{ and }
\pi_1' + N \leftarrow N' + \pi_2' + \pi_2,$$
is included in (\ref{eq:c13BF}). The contributions to $c_1$ produced by the
parts of $P^\dagger$ which are 1PR in
these two channels must be dealt with separately. This, in fact, is the reason
why possibilities 1 and 2 had to be dealt with above. Once this is done,
however, only the part
$\tilde{c}_1$ of $c_1$ is left for consideration. But,
$\tilde{c}_1 \subset c_1 \cap \cfi{N2'}$ if and only if (\ref{eq:c13BF}) is
included in $\cfi{N2'}$. Consequently, in this calculation, in which
$M^{(3)}_1$ is set to zero, the diagrams $\tilde{c}_1$ are not in $c_1 \cap
\cfi{N2'}$.
\end{enumerate}

Turning now to (\ref{eq:appdc2}), similar arguments to the above show that the
expression (\ref{eq:appdc2}) is only part of $\cfi{N2'}$ if a pure three-body
force is included in the calculation. Therefore, for the purposes of this
argument (\ref{eq:appdc2}) is {\em not} in $\cfi{N2'} \cap c_1$.

\section {Details of the calculation of $\lowercase{c}_2 \cap C_5^{\{N2'\}}$}

\label {ap-calcdet2}

The first internal cut lemma may be used to show that the diagrams which
contribute to $\titnni{2}$ and are 2PR in the
$N1' \leftarrow N1 + N2 + N2'$-channel sum to give:
\be
\fir{1}(2) d_1 d_\pi \tilde{F}^{(2)^\dagger},
\ee  where $\tilde{F}^{(2) \dagger}$ is two-particle irreducible in the
$s$-channel and the channel:
$$\pi' + N1' \leftarrow N1 + N2 + N2',$$  as well as being 1PI in the channel:
$$N2' + N2 \leftarrow N1' + \pi' + N1.$$

The question now is, which portions of the two expressions:
\be
\fir{1}_{\pi_1}(1) d_2 d_{\pi_1}  \fir{1}_{\pi_2}(2) d_2  d_{\pi_2}
\tilde{F}^{(2)^\dagger} d_2 \fsai{1}{\pi_1}(1)
\label{eq:app2dc1}
\ee and:
\be
\fir{1}_{\pi_1}(1) d_2 d_{\pi_1} \fir{1}_{\pi_2}(2) d_2  d_{\pi_2}
\tilde{F}^{(2)^\dagger} d_1 d_2 T_{NN}^{(1)} d_2
\fsai{1}{\pi_1}(1)
\label{eq:app2dc2}
\ee
are also in $\cfi{N2'}$? As in Appendix~\ref{ap-calcdet1} we argue that we
are not interested in those portions of expressions (\ref{eq:app2dc1}) and
(\ref{eq:app2dc2}) which may be reexpressed in the form (\ref{eq:c13BF}).
This line of reasoning leads to the conclusion that the diagrams contained in
(\ref{eq:app2dc1}) and $\cfi{N2'}$ are:
\be
\fir{1}_{\pi_1}(2) d_{\pi_1} \fir{1}_{\pi_2}(1) d_2 t_{\pi_1 N}^{(1)}(1) d_2
d_{\pi_2} \fsai{1}{\pi_2}(1) d_1 d_{\pi_1}
\fsai{1}{\pi_1}(2),
\ee  and:
\be
\fir{1}_{\pi_1}(2) d_1 d_{\pi_1} \fir{1}_{\pi_2}(1) d_{\pi_2} d_1
\titnni{1} d_1 \fsai{1}{\pi_2}(1) d_2 d_{\pi_1}
\fsai{1}{\pi_1}(2),
\ee  where $\titnni{1}$ is not only one-particle irreducible in the
$s$-channel, but also 1PI in the $t$-channel.

Similarly, there are no diagrams which are in both (\ref{eq:app2dc2}) and
$\cfi{N2'}$ provided, once again, that no pure three-body force is included in
the calculation.

\begin{figure}
\vspace{35 mm}
\hskip 20 mm
\caption {The term on the right is included in the current $NN-\pi NN$
equations, while the term on the left, known as the Jennings term, is not
included.}
\vskip 4 mm
\label {fig-Jennings}
\end{figure}

\begin{figure}
\vspace{45 mm}
\hskip 40 mm
\vskip 1 mm
\caption {One covariant perturbation theory diagram which is double-counted if
the original Taylor method is used to construct an equation for $\tnni{2}$,
with the two cuts which place it in both
$C_3^{\{N1\}}$ and $C_3^{\{N2\}}$.}
\label{fig-C3dc}
\end{figure}

\begin{figure}
\vspace{35 mm}
\hskip 12 mm
\vskip 3 mm
\caption {If the two cuts shown are three-cuts then they will place certain
 diagrams which should only be included in $C_4$ in $C_5$ as well.}
\label{fig-C4cuts}
\end{figure}

\vfill
\eject
\null

\begin{figure}
\vspace{30 mm}
\hskip 40 mm
\vskip 2  mm
\caption {One possible source of double-counting: if $t_{\pi N}^{(2)}$ is
one-particle reducible in the $u$-channel then the cut shown places this
diagram in $C_4^{\{\pi'\}\{N2\}}$. Since the diagram  arose as part of
$C_4^{\{N2'\}\{N1\}}$ such $u$-channel one-particle reducibility leads to
double-counting.}
\label{fig-dcviau1pr}
\end{figure}

\begin{figure}
\vspace{43 mm}
\hskip 40 mm
\vskip 2 mm
\caption {Another possible source of double-counting: if $t_{\pi N}^{(2)}$
is one-particle reducible in the $t$-channel then the cut shown places this
diagram
in $C_4^{\{N2'\}\{N1\}}$. Since the diagram  arose as part of sub-class
$C_4^{\{N1'\}\{N2\}}$ such $t$-channel one-particle reducibility would lead to
double-counting.}
\label{fig-dcviat1pr}
\end{figure}

\begin{figure}
\vspace{40 mm}
\hskip 15 mm
\vskip 3 mm
\caption {An example of a diagram which would lead to double-counting of the
type depicted in Figure \protect{\ref{fig-dcviat1pr}}.}
\label{fig-pipiia}
\end{figure}

\vfill
\eject
\null

\begin{figure}
\vspace{55 mm}
\hskip 30 mm
\vskip 3 mm
\caption {The first term in the sub-class $C_5^{\{N1'\}}$, $c_1$, with the cut
which may lead to double-counting indicated.}
\label{fig-firstterm}
\end{figure}

\begin{figure}
\vspace{25 mm}
\hskip 27 mm
\vskip 2 mm
\caption {The lowest order diagrams which contribute to $\pi-\pi$ scattering in
a Lagrangian with a $\pi NN$ vertex and a $\pi N$ contact term.}
\label{fig-box}
\end{figure}

\begin{figure}
\vspace{40 mm}
\hskip 27 mm
\vskip 3 mm
\caption {The second term of $C_5^{\{N1'\}}$, $c_2$, with the cut which may
lead to double-counting indicated.}
\label{fig-secondterm}
\end{figure}

\vfill
\eject
\null

\begin{figure}
\vspace{40 mm}
\hskip 30 mm
\vskip 3 mm
\caption {The third term in the sub-class $C_5^{\{N1'\}}$, $c_3$, with  the
cuts which may lead to double-counting indicated.}
\label{fig-term3}
\end{figure}

\begin{figure}
\vspace{45 mm}
\hskip 25 mm
\vskip 3 mm
\caption {The fourth term in $C_5^{\{N1'\}}$, $c_4$, and the cut which may lead
to double-counting.}
\label{fig-Fourthtermcut}
\end{figure}

\begin{figure}
\vspace{45 mm}
\hskip 15 mm
\vskip 2 mm
\caption {The two possible cuts which may lead to double-counting fifth term of
$C_5^{\{N1'\}}$, $c_5$.}
\label{fig-fifthterm}
\end{figure}

\vfill
\eject
\null

\begin{figure}
\vspace{45 mm}
\hskip 15 mm
\vskip 2 mm
\caption {The two possible cuts which may lead to double-counting in the sixth
term of $C_5^{\{N2'\}}$, $c_6$.}
\label{fig-sixthterm}
\end{figure}

\begin{figure}
\vspace{45 mm}
\hskip -8 mm
\vskip 3 mm
\caption {The definition of the new vertex, $f^{(1) *}$.}
\label{fig-Newvertex}
\end{figure}

\vfill
\eject
\null

\begin{figure}
\vspace{75 mm}
\hskip 5 mm
\vskip 3 mm
\caption {Two diagrams which are part of $\tnni{2}$. They will both contribute
to the dressing of the vertices in one-pion exchange when the one-pion exchange
piece of $\tnni{1}$ is inserted, as shown here. Hence both diagrams will lead
to double-counting, unless the vertex used in the one-pion exchange piece of
$\tnni{2}$ is adjusted.}
\label{fig-Vertexovercount}
\end{figure}

\vfill
\eject
\null

\begin{figure}
\vspace{110 mm}
\hskip 1 mm
\vskip 3 mm
\caption {On the right we show the three terms $D_1$, $D_2$ and $D_\pi$. On the
left we depict, in each case, the two different places they appear in the pion
absorption/production piece of $T_{NN}^{(2)}$.}
\label{fig-Vbar2}
\end{figure}

\vfill
\eject
\null

\begin{figure}
\vspace{75 mm}
\hskip 5 mm
\vskip 3 mm
\caption {In the lower half of this diagram we show the two terms $B$ and
$X$, while the upper half displays the two places they appear in the pion
absorption/production piece of $T_{NN}^{(2)}$.}
\label{fig-Vbar1}
\end{figure}

\begin{figure}
\vspace{35 mm}
\hskip 1 mm
\vskip 2 mm
\caption {A diagram containing the input $NN$ amplitude $t_{NN}^{(1)}$. Note
that if the one-pion exchange piece of the $NN$ amplitude is considered this
diagram yields the crossed two-pion exchange graph.}
\label{fig-XedTPE}
\end{figure}

\vfill
\eject
\null

\begin{figure}
\vspace{35 mm}
\hskip 10 mm
\vskip 2 mm
\caption {A piece of an arbitrary Feynman diagram to be included in the
four-dimensional $NN-\pi NN$ equations.}
\label{fig-FDpiece}
\end{figure}

\begin{figure}
\vspace{35 mm}
\hskip 54 mm
\vskip 2 mm
\caption {An example of a diagram in which a self-energy contribution on one of
the legs belonging to the initial state may be isolated.}
\label{fig-Selfenergy}
\end{figure}

\end {document}